\documentclass[draftcls,onecolumn,10pt]{IEEEtran}

\usepackage[T1]{fontenc}
\usepackage[latin1]{inputenc}
\usepackage{cite}      
\usepackage{graphicx}  
\usepackage{psfrag}    
\usepackage{subfigure} 
\usepackage{url}       
\usepackage{amssymb,amsmath,amscd,amsfonts,amstext}
\usepackage{array}
\usepackage{multirow}
\newtheorem{theorem}{Theorem}
\newtheorem{remark}{Remark}
\newtheorem{assumption}{{\bf Assumption}}

\newtheorem{proposition}{Proposition}

\makeatletter
\relax 
\@setckpt{somecommands}{
\setcounter{page}{1}
\setcounter{equation}{0}
\setcounter{enumi}{0}
\setcounter{enumii}{0}
\setcounter{enumiii}{0}
\setcounter{enumiv}{0}
\setcounter{footnote}{0}
\setcounter{mpfootnote}{0}
\setcounter{section}{0}
\setcounter{subsection}{0}
\setcounter{subsubsection}{0}
\setcounter{paragraph}{0}
\setcounter{IEEEsubequation}{0}
\setcounter{figure}{0}
\setcounter{table}{0}
\setcounter{biography}{0}
\setcounter{subfigure}{0}
\setcounter{lofdepth}{1}
\setcounter{subtable}{0}
\setcounter{lotdepth}{1}
\setcounter{parentequation}{0}
\setcounter{theorem}{0}
\setcounter{remark}{0}
\setcounter{assumption}{0}
\setcounter{proposition}{0}
}
\makeatother

\IEEEoverridecommandlockouts 

\begin{document}

\title{BER and Outage Probability Approximations for LMMSE Detectors on Correlated MIMO Channels } 

\author{Abla Kammoun$^{(1)}$\thanks{$(1)$ ENST, Paris, France.
\texttt{abla.kammoun@enst.fr}}, 
Malika Kharouf$^{(2)}$\thanks{$(2)$ Casablanca University, Morocco and ENST, Paris. 
\texttt{malika.kharouf@enst.fr}}, 
Walid Hachem$^{(3)}$ and Jamal Najim$^{(3)}$\thanks{$(3)$ CNRS / ENST, Paris,
France. \texttt{walid.hachem, jamal.najim@enst.fr} }
}


\maketitle

\begin{abstract}

  This paper is devoted to the study of the performance of
  the Linear Minimum Mean-Square Error receiver for (receive)
  correlated Multiple-Input Multiple-Output systems. By the 
  random matrix theory, it is well-known that the Signal-to-Noise Ratio 
  (SNR) at the output of this receiver behaves asymptotically like a 
  Gaussian random variable as the number of receive and transmit antennas 
  converge to $+\infty$ at the same rate. However, this approximation being
  inaccurate for the estimation of some performance metrics such as
  the Bit Error Rate and the outage probability, especially for
  small system dimensions, Li {\it et al.} proposed convincingly to
  assume that the SNR follows a generalized Gamma distribution which 
  parameters are tuned by computing the first three asymptotic moments of 
  the SNR. In this article, this technique is generalized to
  (receive) correlated channels, and closed-form expressions for the first 
  three asymptotic moments of the SNR are provided. 
  To obtain these results, a random matrix theory technique adapted to 
  matrices with Gaussian elements is used. This technique is believed to be 
  simple, efficient, and of broad interest in wireless communications. 
  Simulations are provided, and show that the proposed
  technique yields in general a good accuracy, even for small system
  dimensions.
\end{abstract}

  {\bf Index Terms:} Large random matrices, correlated channels,
  outage probability, Bit Error Rate (BER), Gamma approximation,
  minimum mean square error, Multiple-Input Multiple-Output (MIMO)
  systems, Signal-to-Noise Ratio (SNR).

\section{Introduction}

Since the mid-nineties, digital communications over Multiple Input 
Multiple Output (MIMO) wireless channels have aroused an intense research 
effort. It is indeed well-known since Telatar's work \cite{telatar} that 
antenna diversity increases significantly the Shannon mutual information of a 
wireless link; In rich scattering environments, this mutual information 
increases linearly with the minimum number of transmit and receive antennas. 
Since the findings of \cite{telatar}, a major effort has been devoted to 
analyse the statistics of the mutual information. Such an analysis has strong
practical impacts: For instance, it can provide information about the
gain obtained from scheduling strategies \cite{hochwald}; it can be used as 
a performance metric to optimally select the active transmit antennas 
\cite{ravi}, etc. \\
The early results on MIMO channels mutual information concerned channels
with centered independent and identically distributed entries. 
It is of interest to study the statistics of this mutual information 
for more practical (correlated) MIMO channels. 
In this course, many works established the asymptotic normality of
the mutual information in the large dimension regime
for the so called Kronecker correlated
channels \cite{capacity_analysis,taricco}, for general spatially
correlated channels \cite{moustakas} and for general variance profile 
channels \cite{tlc}.

Another performance index of clear interest is the Signal to Noise
Ratio (SNR) at the output of a given receiver. 
In this paper we focus on one of the most popular receivers, namely
the linear Wiener receiver, also called LMMSE for Linear Minimum Mean
Squared Error receiver. In this context, an \emph{outage} event occurs when
the SNR at the LMMSE output lies beneath a given threshold. 
One purpose of this paper is to approximate the associated outage probability 
for an important class of MIMO channel models. Another performance index 
associated with the SNR is the Bit Error Rate (BER) which will be
also studied herein. \\
Outage probability approximations has been provided in recent works for
various channels, under very specific technical conditions (in the
case where the moment generating function \cite{ko} or the probability
density function \cite{stuber} have closed form expressions; when a
first order expansion of the probability density function can be
derived \cite{Mckay}; in the more general case where the moment
generating function can be approximated by using Pad\'{e}
approximations \cite{stokes}; etc.).
All these results deal with specific situations where the statistics of the 
SNR could be derived for finite system dimensions. \\ 
Alternatively, by making use of large random matrix theory, one can study the 
behavior of the SNR in the asymptotic regime where the channel matrix 
dimensions grow to infinity. For fairly general channel statistical models, 
it is then possible to prove the convergence of the SNR to deterministic 
values and even establish its asymptotic normality (see for instance 
\cite{spawc1,clt_journal}).  
However, this Gaussian approximation is not accurate when the channel
dimensions are small. 
This is confirmed in \emph{e.g.} \cite{guo} where it is shown that the 
asymptotic BER based on the sole Gaussian approximation is significantly 
smaller than the empirical estimate.
A more precise approximation of the BER or the outage probability
is expected if one chooses to approximate the SNR probability distribution
with a distribution 1) which is supported by $\mathbb{R}_+$ 
(indeed, a Gaussian random variable takes negative values which is not 
realistic), 2) which is adjusted to the first three moments of the SNR 
instead of the first two moments needed by the Gaussian approximation. \\ 
In this line of thought, 
Li, Paul, Narasimhan and Cioffi \cite{cioffi} proposed to use
alternative parameterized distributions (Gamma and generalized Gamma
distributions) whose parameters are set to coincide with the
asymptotic moments of the output SNR. This approach was
derived for (transmit) correlated channels and asymptotic moments were
provided for the special case of uncorrelated or equicorrelated channels.
For the general correlated channel case, only limiting upper bounds for the first three asymptotic moments were provided.
Based on
Random Matrix Theory and especially on the Gaussian mathematical tools
elaborated in \cite{capacity_analysis} and further used in
\cite{algo}, we derive closed-form expressions for the first three
moments, generalizing the work of \cite{cioffi} to a general (receive)
correlated channel. Using the generalized Gamma approximation, we
provide closed-form expressions for the BER and numerical
approximations for the outage probability.

{\it Paper organization }

In section II, we present the system model and derive the SNR expression. Then we review in section III the Generalized Gamma approximation before providing the asymptotic central moments in the next section. Finally, we discuss in the last section the simulation results.

\section{System Model and SNR expression}\label{system}

We consider an uplink transmission system, in which a base station
equipped by $N$ correlated antennas detects the symbols of a given
user of interest in the presence of $K$ interfering users. The $N$
dimensional received signal writes:
$$
{\bf r}={\bf \Sigma} {\bf s}+{\bf n},
$$
where ${\bf s}=\left[s_0,\cdots,s_K\right]^{\mathrm{\small T}}$ is the
transmitted complex vector signal with size $K+1$ satisfying
$\mathbb{E}{\bf ss}^*={\bf I}_{K+1}$, and ${\bf \Sigma}$ is the
$N\times (K+1)$ channel matrix. We assume that this matrix writes as 
$$
{\bf \Sigma}=\frac{1}{\sqrt{K}}\boldsymbol{\Psi}^{\frac{1}{2}}{\bf W}{{\bf P}}^{\frac{1}{2}},
$$
where $\boldsymbol{\Psi}$ a $N\times N$ Hermitian nonnegative matrix that
captures the correlations at the receiver, 
${\bf P}=\mathrm{diag}\left(p_0,\cdots,p_K\right)$ is the
deterministic matrix of the powers allocated to the different users
and ${\bf W}=\left[{\bf w}_0,\cdots,{\bf w}_K\right]$ (${\bf w}_k$
being the $k$th column) is a $N\times (K+1)$ complex Gaussian matrix
with centered unit variance (standard) 
independent and identically distributed (i.i.d) 
entries. To detect symbol $s_0$ and to mitigate the
interference caused by users $1,\ldots, K$, the base station applies the
LMMSE estimator, which minimizes the following metric:
$$
{\bf g}=\min_{{\bf h}}\mathbb{E}\left|{\bf h}^*{\bf r}-s_0\right|^2 \ . 
$$
Let ${\bf y}=\sqrt{\frac{p_0}{K}}\boldsymbol{\Psi}^{\frac{1}{2}}{\bf w}_0$, 
 then it is well known that the LMMSE estimator is given by:
$$
{\bf g}=\left({\bf \Sigma}{\bf \Sigma}^*+\rho{\bf I}_N\right)^{-1}{\bf y}.
$$
Writing the received vector ${\bf r}=s_0{\bf y}+{\bf r}_{\mathrm{in}}$
where ${s_0{\bf y}}$ is the relevant term and ${\bf r}_{\mathrm{in}}$
represents the interference plus noise term, the SNR at the output of
the LMMSE estimator is given by : $\beta_K=\left|{\bf g}^*{\bf
    y}\right|^2 /\mathbb{E}\left|{\bf g}^*{\bf
    r}_{\mathrm{in}}\right|^2$. Plugging the expression of ${\bf g}$
given above into this expression, one can show that the SNR $\beta_K$
is given by:
$$
\beta_K={\bf y}^* \left( \frac 1K
{\boldsymbol \Psi}^{\frac 12} \widetilde{\bf W} 
\widetilde{\bf P} \widetilde{\bf W}^* {\boldsymbol \Psi}^{\frac 12} 
+\rho {\bf I}_N \right)^{-1}{\bf y},
$$
with 
$\widetilde{\bf P}=\mathrm{diag}\left(p_1,\cdots,p_K\right)$ and 
$\widetilde{\bf W}=\left[{\bf w}_1,\cdots,{\bf w}_K\right]$. 
Let $\boldsymbol{\Psi}={\bf U}\boldsymbol{\bf D} {\bf U}^*$ be a spectral decomposition of $\boldsymbol{\Psi}$. Then, $\beta_K$ writes: 
\begin{eqnarray*}
  \beta_K&=&\frac{{p}_0}{K}{\bf w}_0^*{\bf U}{\bf D}^{\frac{1}{2}}\left(\frac{1}{K}{\bf D}^{\frac{1}{2}}{\bf U}^*\widetilde{\bf W}\widetilde{\bf P}\widetilde{\bf W}^*{\bf U}{\bf D}^{\frac{1}{2}}+\rho {\bf I}_N\right)^{-1}{\bf D}^{\frac{1}{2}}{\bf U}^*{\bf w}_0\ ,\\
  &=&\frac{p_0}{\rho K}{\bf z}^*{\bf D}^{\frac{1}{2}}\left(\frac{1}{K\rho}{\bf D}^{\frac{1}{2}}{\bf Z}\tilde{{\bf D}}{\bf Z}^*{\bf D}^{\frac{1}{2}}+ {\bf I}\right)^{-1}{\bf D}^{\frac{1}{2}}{\bf z}
\end{eqnarray*}
where:
${\bf z}={\bf U}^*{\bf w}_0$ (resp. ${\bf Z}={\bf U}^*\widetilde{\bf
  W}$) is a $N\times 1$ vector with complex independent standard Gaussian
entries (resp. $N\times K $ matrix with independent Gaussian entries).

Under appropriate assumptions, it can be proved that $\beta_K$ admits
a deterministic approximation as $K,N\rightarrow\infty$, the ratio
being bounded below by a positive constant and above by a finite constant.
Furthermore, its fluctuations can be precisely described under the
same asymptotic regime (for a full and rigorous computation based on
random matrix theory, see\cite{clt_journal}). As it will appear
shortly, a deterministic approximation of the third centered moment of
$\beta_K$ is needed and will be computed in the sequel.

\section{Bit Error Rate and outage probability approximations}
\subsection{A quick reminder of the generalised Gamma distribution}
Recall that if a random variable $X$ follows a generalized gamma distribution
$G(\alpha, b,\xi)$, where $\alpha$ and $b$ are respectively referred
to as the shape and scale parameters, then:
$$
\mathbb{E}X=\alpha b,\quad \mathrm{var}(X)=\alpha b^2
\quad \textrm{and} \quad \mathbb{E}(X-\mathbb{E}X)^3=(\xi+1)\alpha b^3\ .
$$
The probability density function (pdf) of the generalized Gamma
distribution with parameters ($\alpha,b,\xi$) does not have a closed
form expression but its moment generating function (MGF) writes:
$$
\mathrm{MGF}(s)=\left\{\begin{array}{lc}
\exp(\frac{\alpha}{\xi-1}(1-(1-b\xi s)^{\frac{\xi-1}{\xi}})) \hspace{0.2cm} & \mathrm{if}\ \xi > 1, \\
\exp(\frac{\alpha}{1-\xi}((1-b\xi s)^{\frac{\xi-1}{\xi}}-1)) \hspace{0.2cm} & \mathrm{if} \ \xi \le 1. \\
\end{array}\right.
$$

\subsection{BER approximation}
Under QPSK constellations with Gray encoding and assuming that the
noise at the LMMSE output is Gaussian, the BER is given by:
$$
\mathrm{BER}=\mathbb{E}Q(\sqrt{\beta_K})
$$
where $Q(x)=\frac{1}{\sqrt{2\pi}}\int_{x}^{\infty}e^{-t^2/2}\,dt$ and
the expectation is taken over the distribution of the SNR $\beta_K$.
Based on the asymptotic normality of the SNR, \cite{zeitouni} and
\cite{verdu_ber} proposed to use the limiting $\mathrm{BER}$ value
given by:
$$
\mathrm{BER}=\frac{1}{\sqrt{2\pi}}\int_{\sqrt{\overline{\beta}_K}}^{\infty}e^{-t^2/2}dt,
$$
where $\overline{\beta}_K$ denotes an asymptotic deterministic
approximation of the first moment of $\beta_K$. It was shown however
in \cite{cioffi} that this expression is inaccurate since a Gaussian
random variable allows negative values and has a zero third moment
while the output SNR is always positive and has a non-zero
third moment for finite system dimensions.  To overcome these
difficulties, Li {\it et al.} \cite{cioffi} approximate the BER by
considering first that the SNR follows a Gamma distribution with scale
$\alpha$ and shape $b$, these parameters being tuned by equating the
first two moments of the Gamma distribution with the first two
asymptotic moments of the SNR. However, the third asymptotic moment
was shown to be different from the third moment of the Gamma
distribution which only depends on the scale $\alpha$ and shape $b$.
In light of this consideration, Li {\it et al.} \cite{cioffi} refine
this approximation and consider that the SNR follows a generalized
Gamma distribution which is adjusted by assuming that its first three
moments equate the first three asymptotic moments of the SNR. As
expected, this approximation has proved to be more accurate than
the Gamma approximation, and so will be the one considered in this
paper. Next, we briefly review this technique, which we will rely on
to provide accurate approximations for the BER and outage probability.

Let $\mathbb{E}_{\infty}(\beta_K)$, $\mathrm{var}_{\infty}(\beta_K)$
and $\mathrm{S}_{\infty}(\beta_K)$ denote respectively the
deterministic approximations of the asymptotic central moments of
$\beta_K$. Then, the parameters $\xi$, $\alpha$ and $b$ are determined
by solving:
$$
\mathbb{E}_{\infty}(\beta_K)=\alpha b, \quad  \mathrm{var}_{\infty}(\beta_K)=\alpha b^2 \quad 
\mbox{and}\quad \mathrm{S}_{\infty}(\beta_K)=(\xi+1)\alpha b^3,
$$
thus giving the following values:
$$
\alpha=\frac{(\mathbb{E}_{\infty}(\beta_K))^2}{\mathrm{var}_{\infty}(\beta_K)},
\quad
\beta=\frac{\mathrm{var}_{\infty}(\beta_K)}{\mathbb{E}_{\infty}(\beta_K)}
\quad \mbox{and}\quad
\xi=\frac{\mathrm{S}_{\infty}(\beta_K)\mathbb{E}_{\infty}(\beta_K)}{(\mathrm{var}_{\infty}(\beta_K))^2}-1.
$$
Using the MGF, one can evaluate the BER by using the following
relation \cite{alouini}, that holds for QPSK constellation:
\begin{equation}
\mathrm{BER}=\frac{1}{\pi}\int_0^{\frac{\pi}{2}}\mathrm{MGF}\left(-\frac{1}{2\sin^2\phi}\right)d\phi.
\label{eq:ber}
\end{equation}
Note that similar expressions for the BER exist for other
constellations and can be derived by plugging the following identity
involving the function $Q(x)$ \cite{alouini}:
$$
Q(x)=\frac{1}{\pi}\int_0^{\frac{\pi}{2}}\exp\left(-\frac{x^2}{2\sin^2\theta}\right)d\theta
$$
into the BER expression.
\subsection{Outage probability approximation}
Only the moment generation function (MGF) has a closed form
expression. Knowing the MGF, one can compute numerically the
cumulative distribution function by applying the saddle point
approximation technique \cite{saddle}. Denote by
$K(y)=\log(\mathrm{MGF}(y))$ the cumulative generating function, by
$y$ the threshold SNR and by $t_y$ the solution of $K'(t_y)=y$. Let
$w_0$ and $u_0$ be given by: $ w_0=\mathrm{sign}(t_y)\sqrt{2\left(t_y
    y-K(t_y)\right)}$ and $u_0 = t_y\sqrt{K"(t_y)}$. The saddle
point approximate of the outage probability is given by:
\begin{equation}
\mathrm{P}_{out}=\Phi(w_0)+\phi(w_0)\left(\frac{1}{w_0}-\frac{1}{u_0}\right),
\label{eq:saddle}
\end{equation}
where $\Phi(x)=\int_{-\infty}^x\frac{1}{\sqrt{2\pi}}e^{-t^2/2}\, dt$ and
$\phi(x)=\frac{1}{\sqrt{2\pi}}e^{-x^2/2}$ denote respectively the
standard normal cumulative distribution function and probability
distribution function.

So far, we have presented the technique that will be used in
simulations for the evaluation of the BER and outage probability. This
technique is heavily based on the computation of the three first
asymptotic moments of the SNR $\beta_K$, an issue that is handled in
the next section.
\section{Asymptotic moments}
\subsection{Assumptions}
Recall from Section \ref{system} the various definitions $K,N,{\bf D},\widetilde{\bf D}$.
In the following, we assume that both $K$ and $N$ go to $+\infty$, their ratio being
bounded below and above as follows:
$$
0\  <\  \ell^- =\liminf \frac KN \  \le \  \ell^+ =\limsup \frac KN \  < \  +\infty\ .
$$
In the sequel, the notation $K\to\infty$ will refer to this asymptotic
regime. We will frequently write ${\bf D}_K$ and $\widetilde{\bf D}_K$
to emphasize the dependence in $K$, but may drop the subscript $K$ as
well. Assume the following mild conditions:
\begin{assumption}
\label{ass:first}
There exist real numbers $d_{\mathrm{max}}<\infty$ and  $\tilde{d}_{\mathrm{max}}<\infty$ such that:
$$
\sup_K\|{\bf D}_K\|\leq d_{\mathrm{max}} \hspace{0.5cm} \mathrm{and} Â \hspace{0.5cm} \sup_K\|\widetilde{\bf D}_K\| \leq \tilde{d}_{\mathrm{max}},
$$
where $\|{\bf D}_K\|$ and $\|\widetilde{\bf D}_K\| $ are the spectral norms of 
${\bf D}_K$ and $\widetilde{\bf D}_K$.
\end{assumption}
\begin{assumption}
\label{ass:second}
The normalized traces of ${\bf D}_K$ and $\widetilde{\bf D}_K$ satisfy:
$$
\inf_K\frac{1}{K}\mathrm{Tr}({\bf D}_K) > 0 \hspace{0.5cm} \mathrm{and} \hspace{0.5cm}
\inf_K \frac{1}{K}\mathrm{Tr}(\widetilde{\bf D}_K) > 0.
$$
\end{assumption} 
\subsection{Asymptotic moments computation}
In this section, we provide closed form expressions for the first
three asymptotic moments. We shall first introduce some deterministic
quantities that are used for the computation of the first, second and
third asymptotic moments.

\begin{proposition} (cf. \cite{capacity_analysis}) For every integer
  $K$ and any $t>0$, the system of equations in
  $(\delta,\tilde{\delta})$
$$
\left\{
\begin{array}{lll}
\delta_K&=&\frac{1}{K}\mathrm{Tr}{\bf D}_K\left({\bf I}+t\tilde{\delta}_K\bf{D}_K\right)^{-1},\\
\tilde{\delta}_K&=&\frac{1}{K}\mathrm{Tr}\widetilde{\bf D}_K\left({\bf I}+t\delta_K\widetilde{\bf D}_K\right)^{-1},\\
\end{array}
\right.
$$
admits a unique solution
$\left(\delta_K(t),\tilde{\delta}_K(t)\right)$ satisfying $\delta_K(t)
> 0$, $\tilde{\delta}_K(t) > 0$.
\end{proposition}
Let ${\bf T}$ and $\widetilde{\bf T}$ be the $N\times N$ and $K\times K$ diagonal matrices defined by:
$$
{\bf T}=\left({\bf I}+t\tilde{\delta_K}{\bf D}\right)^{-1} \ \ \textnormal{and} \ \ \widetilde{\bf T}=\left({\bf I}+t{\delta_K}\widetilde{\bf D}\right)^{-1}.
$$
Note that in particular: $\delta=\frac{1}{K}\mathrm{Tr}{\bf DT}$ and
$\tilde{\delta}=\frac{1}{K}\mathrm{Tr}\widetilde{\bf D}\widetilde{\bf
  T}$. Define also $\gamma$ and $\tilde{\gamma}$ as
$\gamma=\frac{1}{K}\mathrm{Tr}{\bf D}^2{\bf T}^2$ and
$\tilde{\gamma}=\frac{1}{K}\mathrm{Tr}\widetilde{\bf
  D}^2\widetilde{\bf T}^2$.
Finally, replace $t$ by $\frac 1\rho$ and introduce the following deterministic quantities:
\begin{eqnarray*}
  \Omega_K^2&=&\frac{\gamma}{\rho^2}\left(\frac{\gamma\tilde{\gamma}}{\rho^2-\gamma\tilde{\gamma}}+1\right),\\
  \nu_K&=&\frac{2\rho^3}{K\left(\rho^2-\gamma\tilde{\gamma}\right)^3}
  \left[\mathrm{Tr}{\bf D}^3{\bf T}^3-\frac{\gamma^3}{\rho^3}\mathrm{Tr}\widetilde{\bf D}^3\widetilde{\bf T}^3\right]\ .
\end{eqnarray*}
As usual, the notation $\alpha_K={\mathcal O}(\beta_K)$ means that
$\alpha_K (\beta_K)^{-1}$ is uniformly bounded as $K\rightarrow
\infty$. Then, the first three asymptotic moments are given by the
following theorem:
\begin{theorem}\label{asymptotic}
Assuming that the matrices ${\bf D}$ and $\widetilde{\bf D}$ satisfy the conditions stated in \ref{ass:first} and \ref{ass:second}, then the following convergences hold true:
\begin{enumerate}
\item First asymptotic moment \cite{spawc1,clt_journal}:
$$
\frac{\delta_K}{\rho} = {\mathcal O}(1)\quad \textrm{and} \quad\mathbb{E}\left(\frac{\beta_K}{p_0}\right)-\frac{\delta_K}{\rho}\xrightarrow[K\to\infty]{} 0,
$$
\item Second asymptotic moment \cite{spawc1,clt_journal}:
$$
\Omega_K = {\mathcal O}(1)\quad \textrm{and} \quad
K\mathbb{E}\left(\frac{\beta_K}{p_0}-\mathbb{E}\left(\frac{\beta_K}{p_0}\right)\right)^2-{\Omega_K^2}\xrightarrow[K\to\infty]{}
0,
$$
\item Third asymptotic moment:
$$
\nu_K = {\mathcal O}(1)\quad \textrm{and} \quad
K^2\mathbb{E}\left(\frac{\beta_K}{p_0}-\mathbb{E}\left(\frac{\beta_K}{p_0}\right)\right)^3-{\nu_K}\xrightarrow[K\to\infty]{} 0.
$$
\end{enumerate}
\end{theorem}
The two first items of the theorem are proved in \cite{clt_journal}
(beware that the notations used in this article are the same as those
in \cite{capacity_analysis} and slightly differ from those used in
\cite{clt_journal}). Proof of the third item of the theorem is
postponed to the appendix.

\begin{remark}
  One can note that the third asymptotic moment is of order ${\mathcal
    O}(K^{-2})$. This is in accordance with the asymptotic normality
  of the SNR, where the third moment of $\sqrt{K} (\beta_K
  -\mathbb{E}(\beta_K))$ will eventually vanish, as this quantity
  becomes closer to a Gaussian random variable. However, its value
  remains significant for small dimension systems.
\end{remark}

\section{Simulation results}

\begin{figure*}[htp] 
\centering
\subfigure[ First moment of the SNR]{
\includegraphics[width=0.5\textwidth]{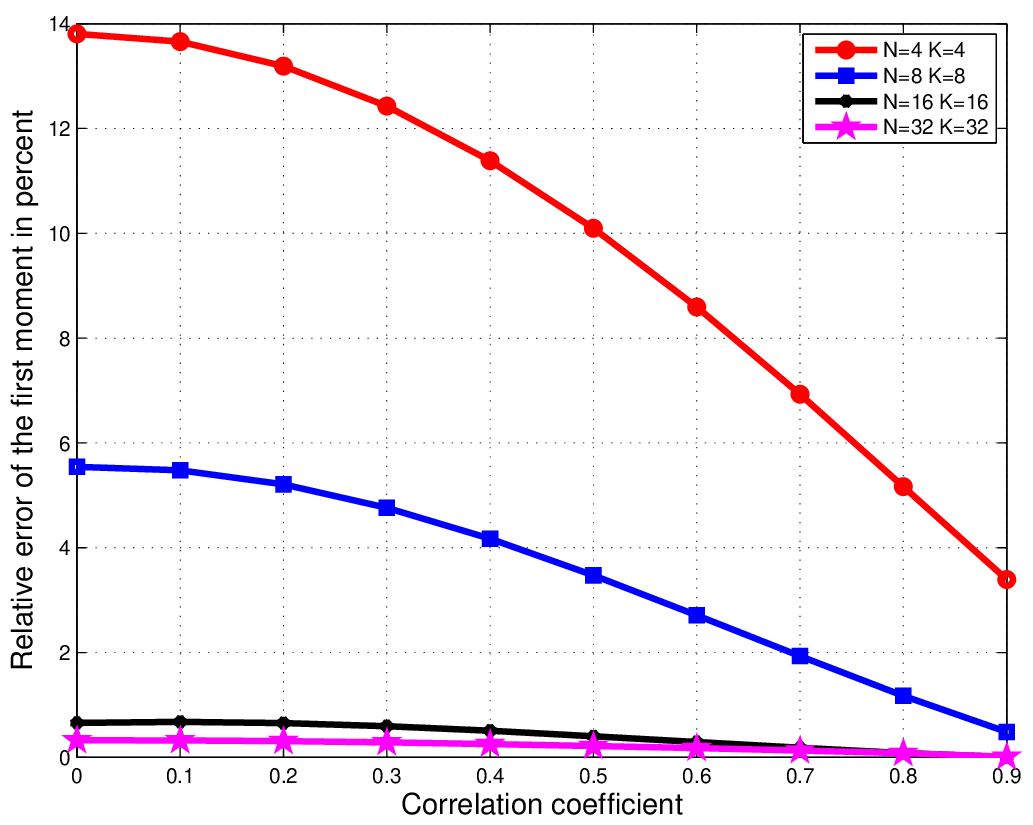}
} 
\subfigure[ Second moment of the SNR]{
\includegraphics[width=0.5\textwidth]{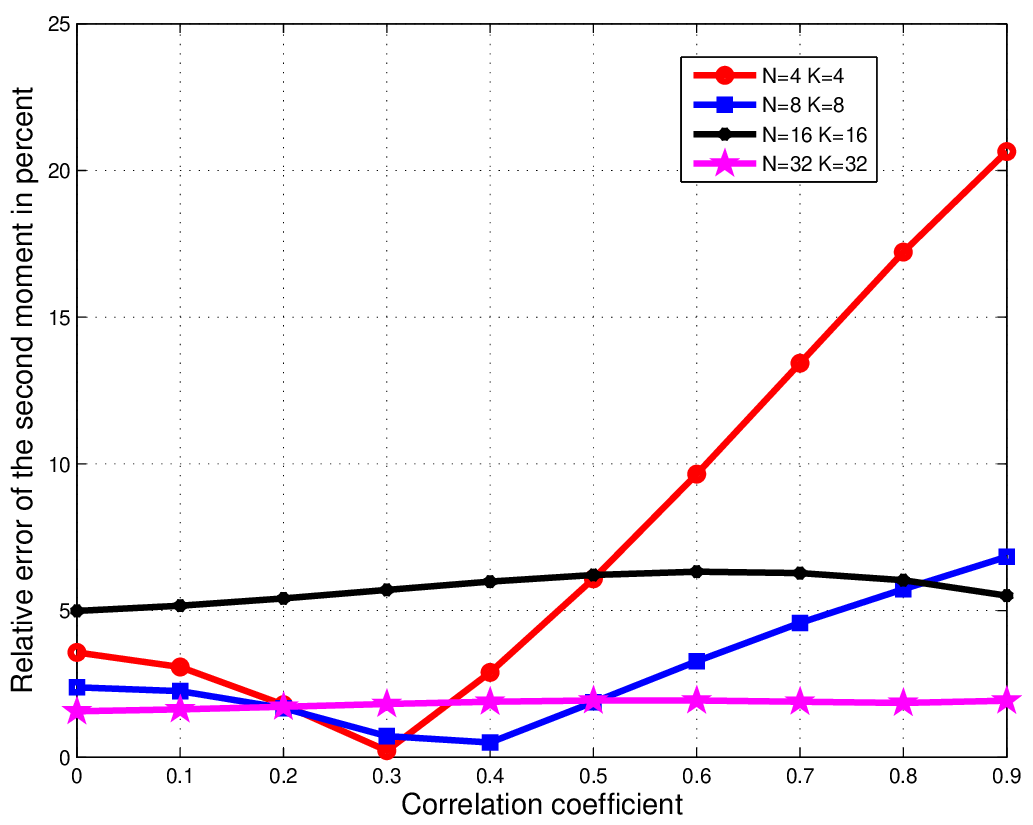}
}
\subfigure[Third moment of the SNR]{
\includegraphics[width=0.5\textwidth]{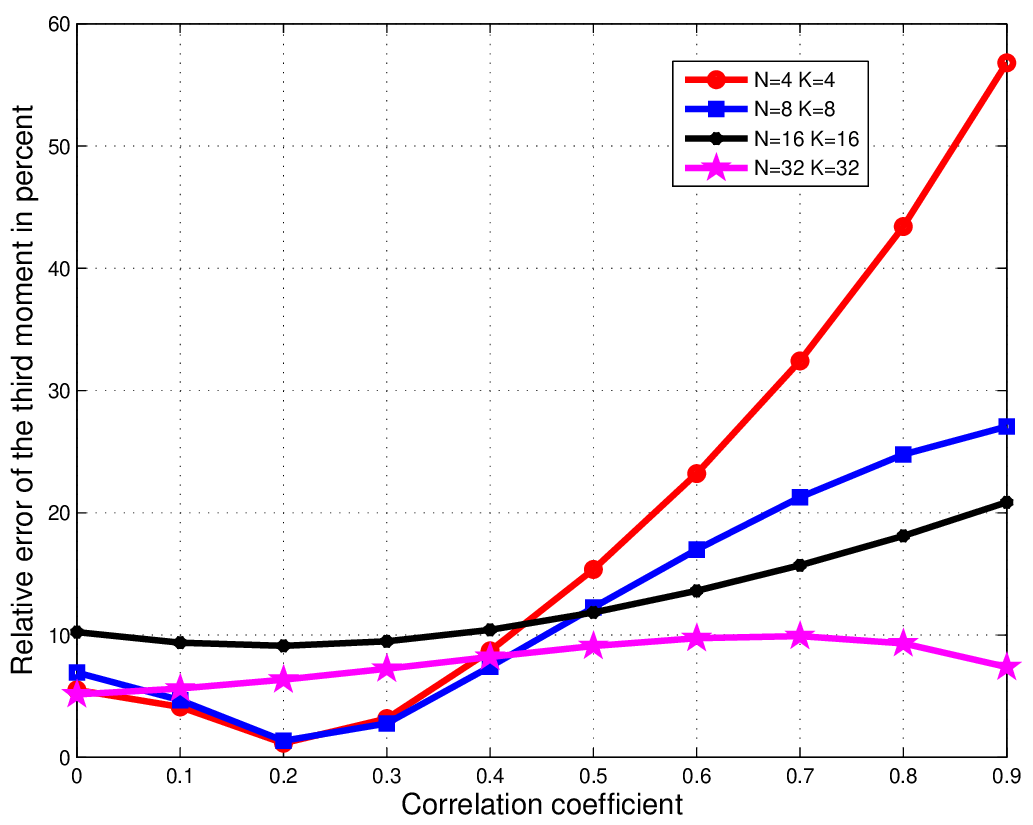}
}
\caption{Absolute value of the relative error when $N=K$ }
\label{fig:rel_err}
\end{figure*}

\begin{figure*}[htp] 
\centering
\subfigure[ First moment of the SNR]{
\includegraphics[width=0.5\textwidth]{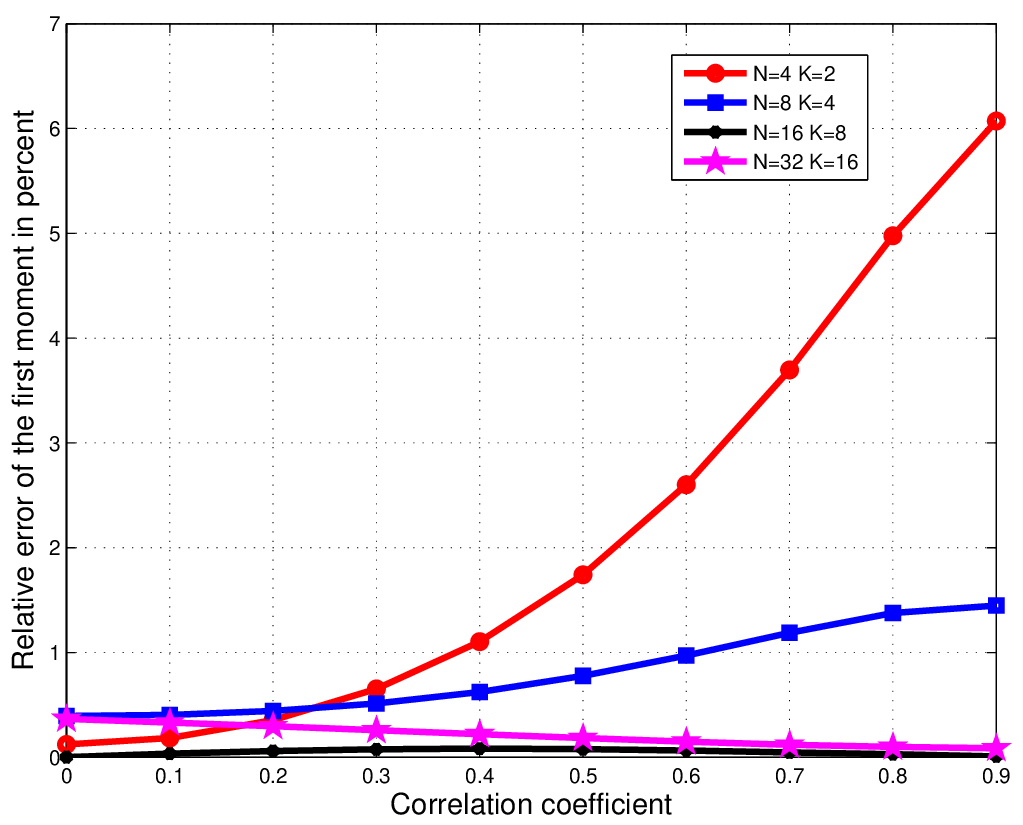}
}
\subfigure[ Second moment of the SNR]{
\includegraphics[width=0.5\textwidth]{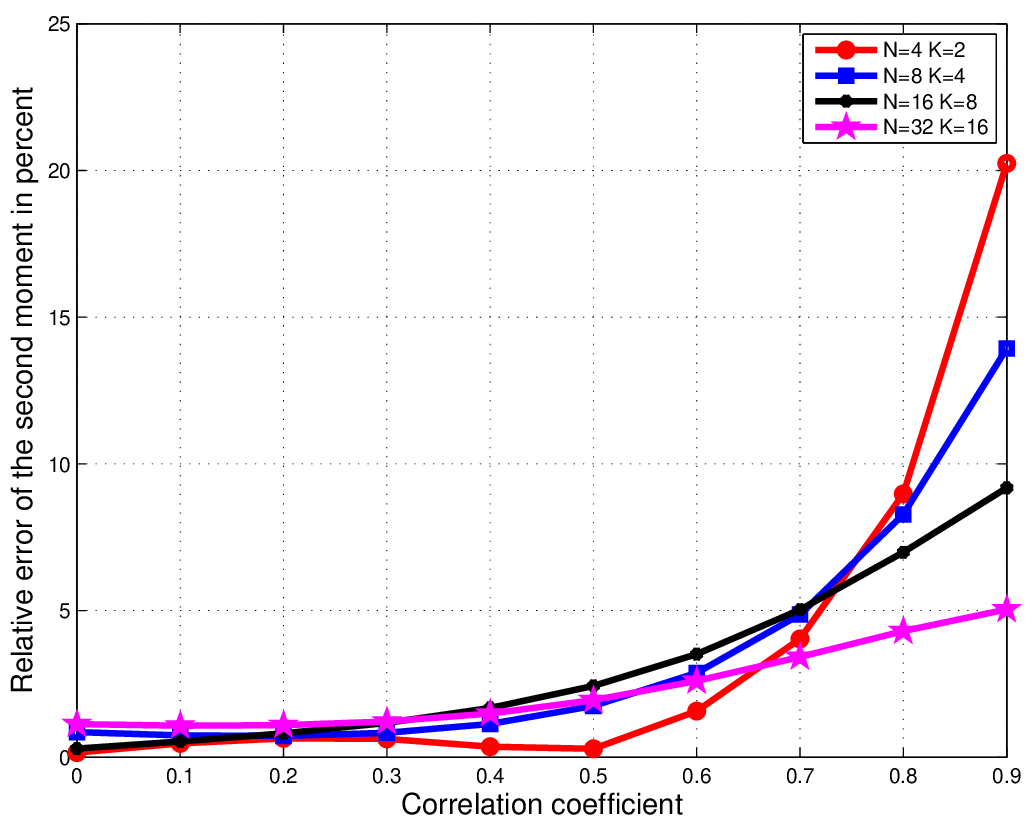}
}
\subfigure[Third moment of the SNR]{
\includegraphics[width=0.5\textwidth]{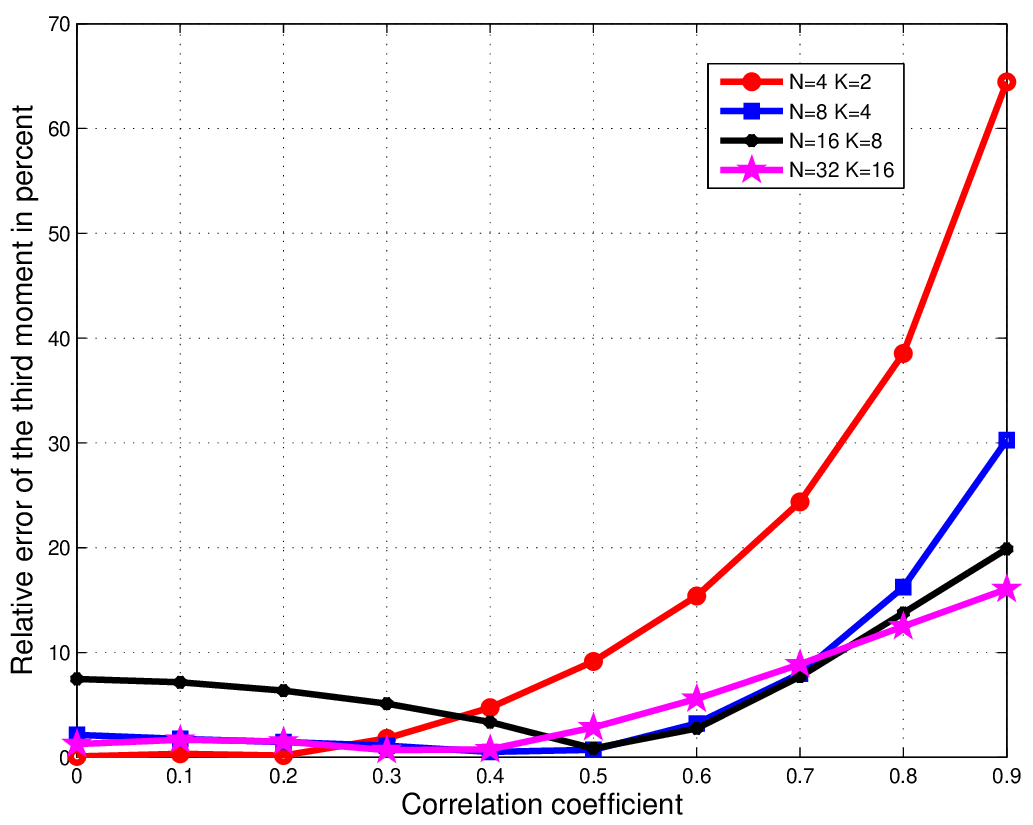}
}
\caption{Absolute value of the relative error when $N=2K$ }
\label{fig:rel_err_neq}
\end{figure*}

\begin{figure*}[htp]
\centering
\subfigure[$N=K=4$ and $a=0$]{
\includegraphics[width=0.4\textwidth]{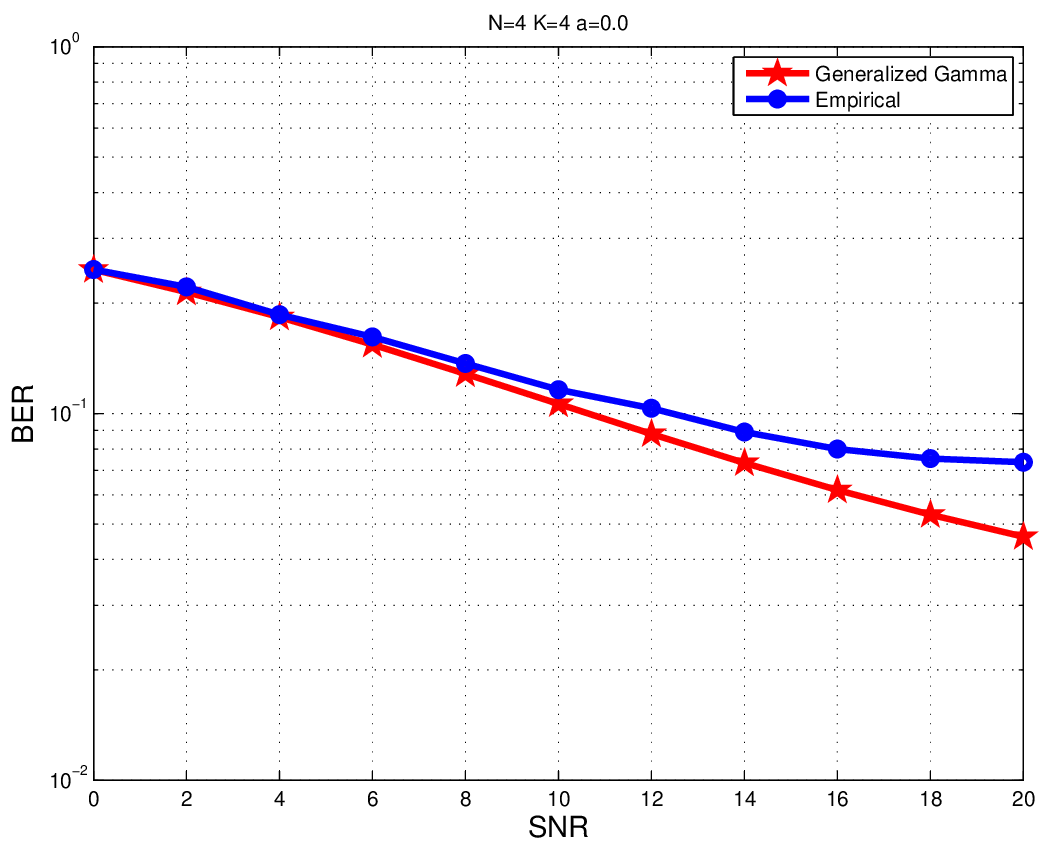}
}
\subfigure[$N=K=4$ and $a=0.9$]{
\includegraphics[width=0.4\textwidth]{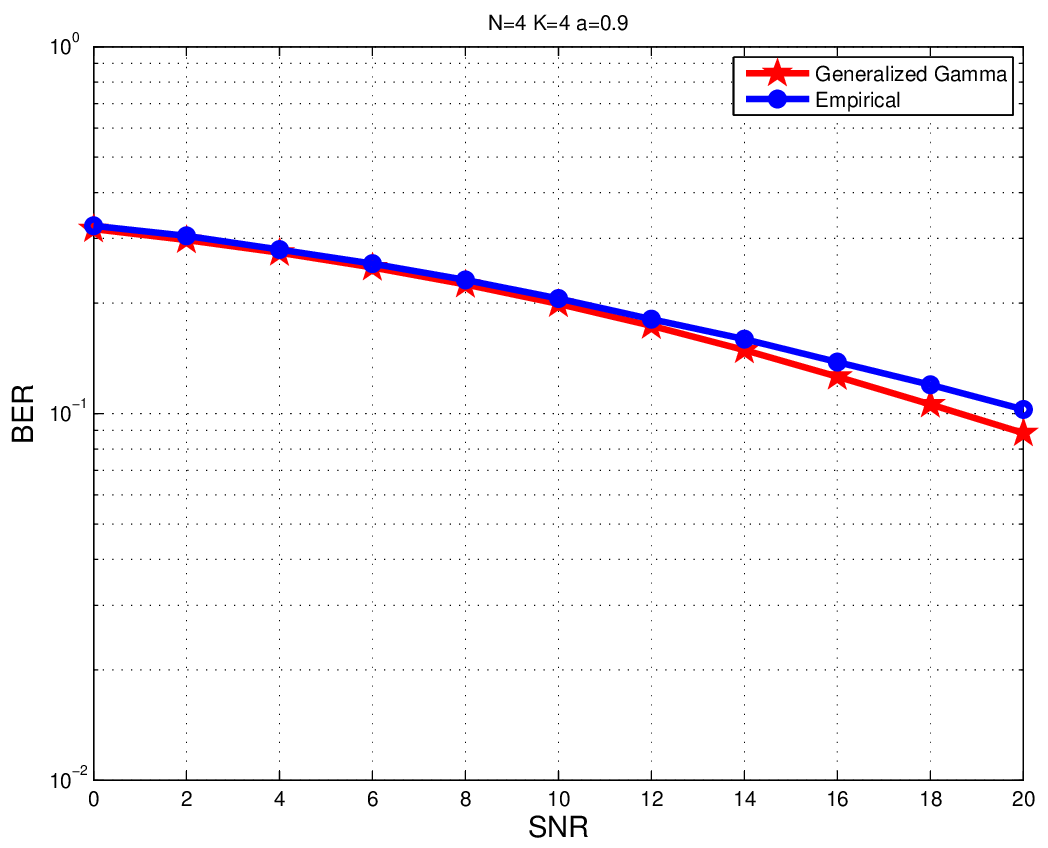}
}
\centering
\subfigure[$N=2K=4$ and $a=0$]{
\includegraphics[width=0.4\textwidth]{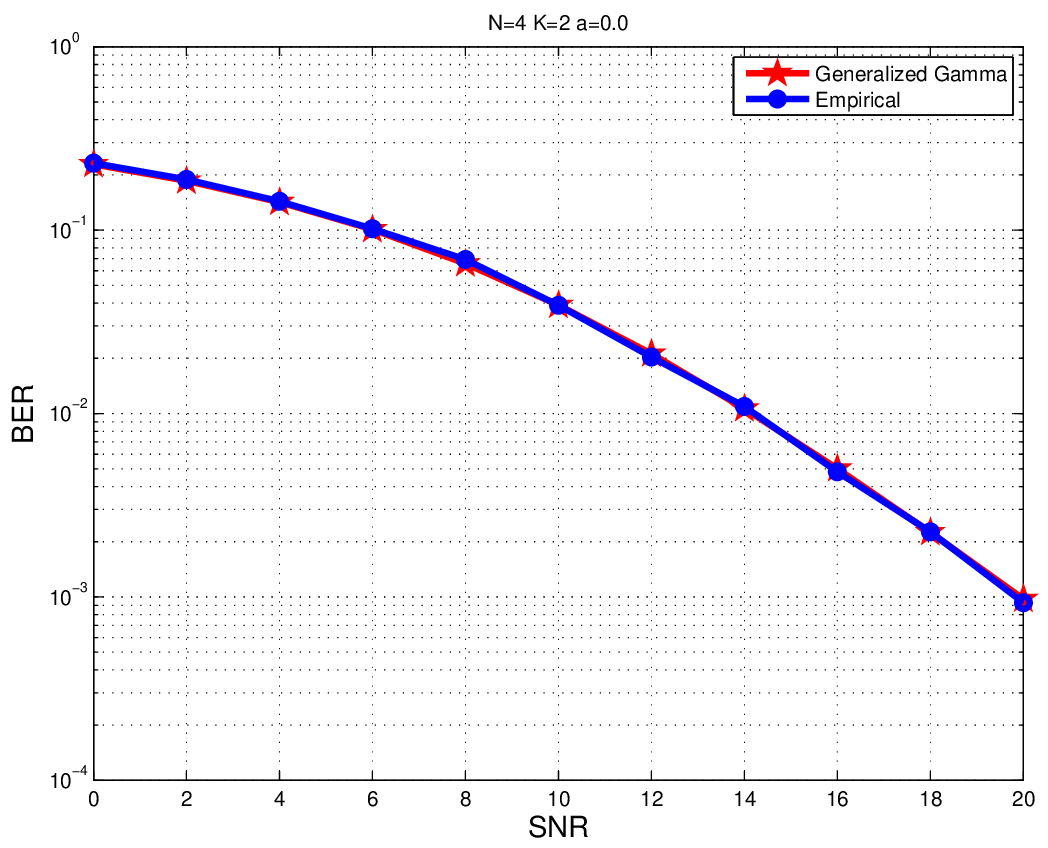}
}
\subfigure[$N=2K=4$ and $a=0.9$]{
\includegraphics[width=0.4\textwidth]{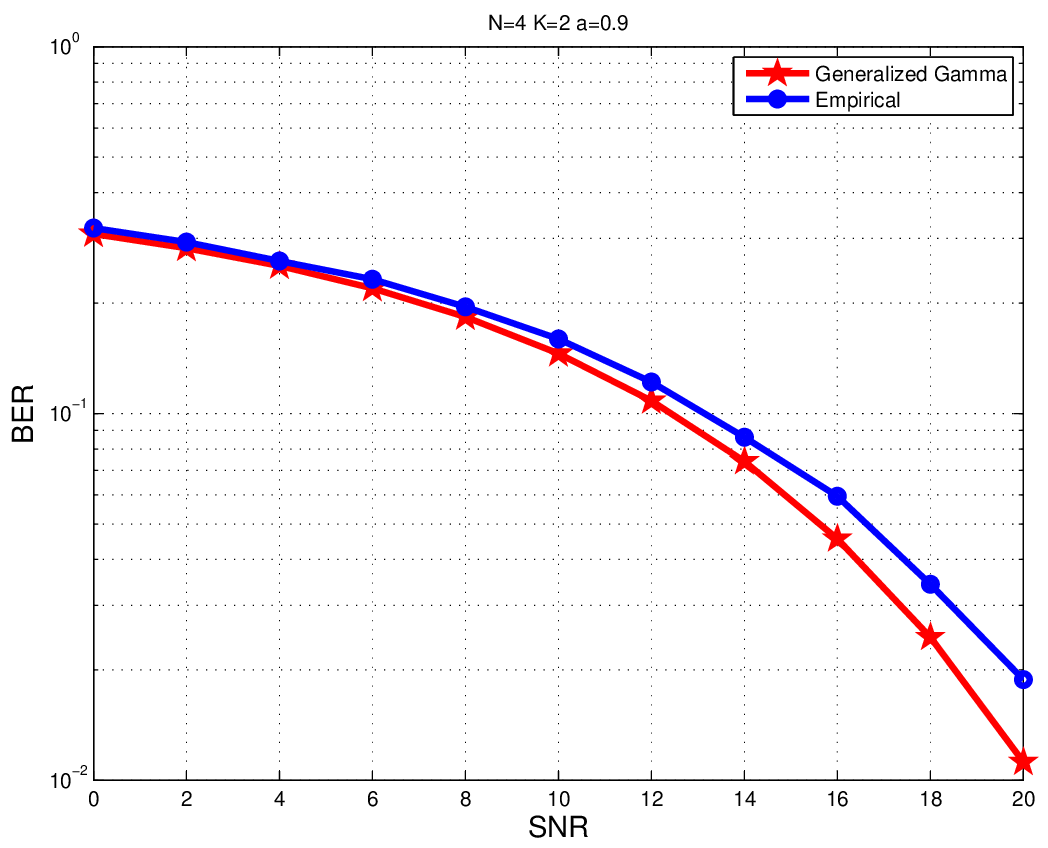}
}
\caption{BER vs input SNR}
\label{fig:BER-vs-snr}
\end{figure*}

\begin{figure*}[htp]
\centering
\subfigure[$N=K=4$ and $a=0$]{
\includegraphics[width=0.4\textwidth]{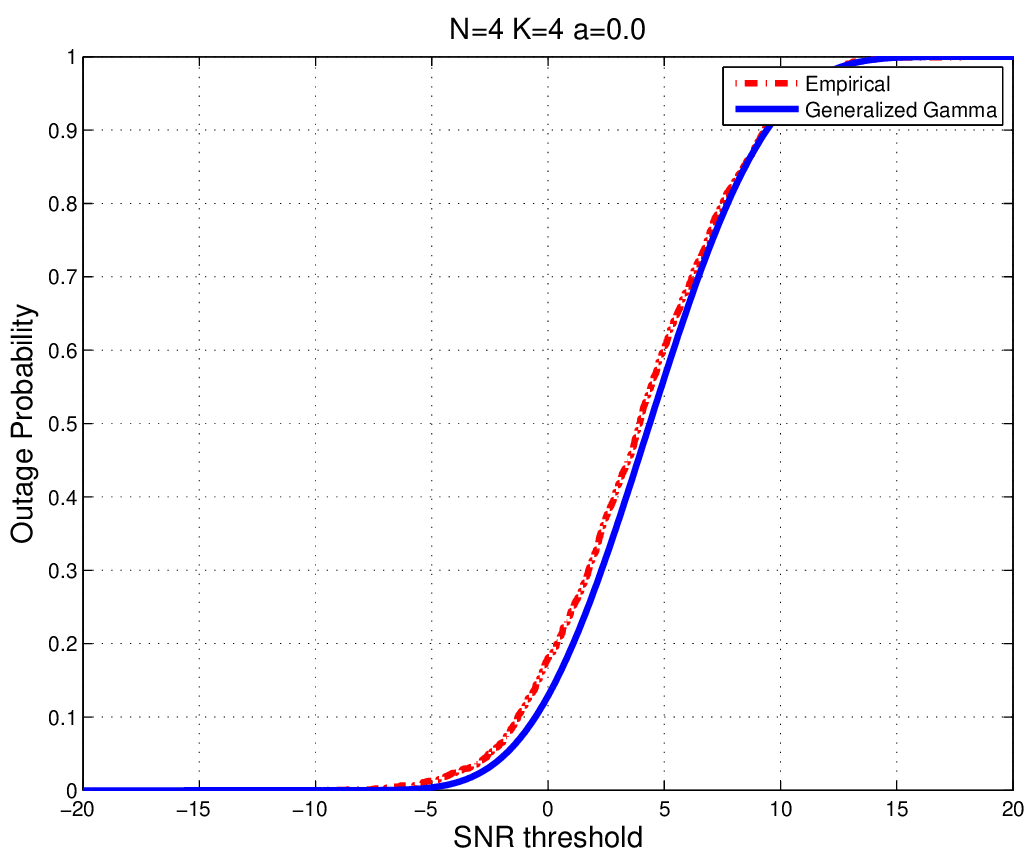}
}
\subfigure[$N=K=4$ and $a=0.9$]{
\includegraphics[width=0.4\textwidth]{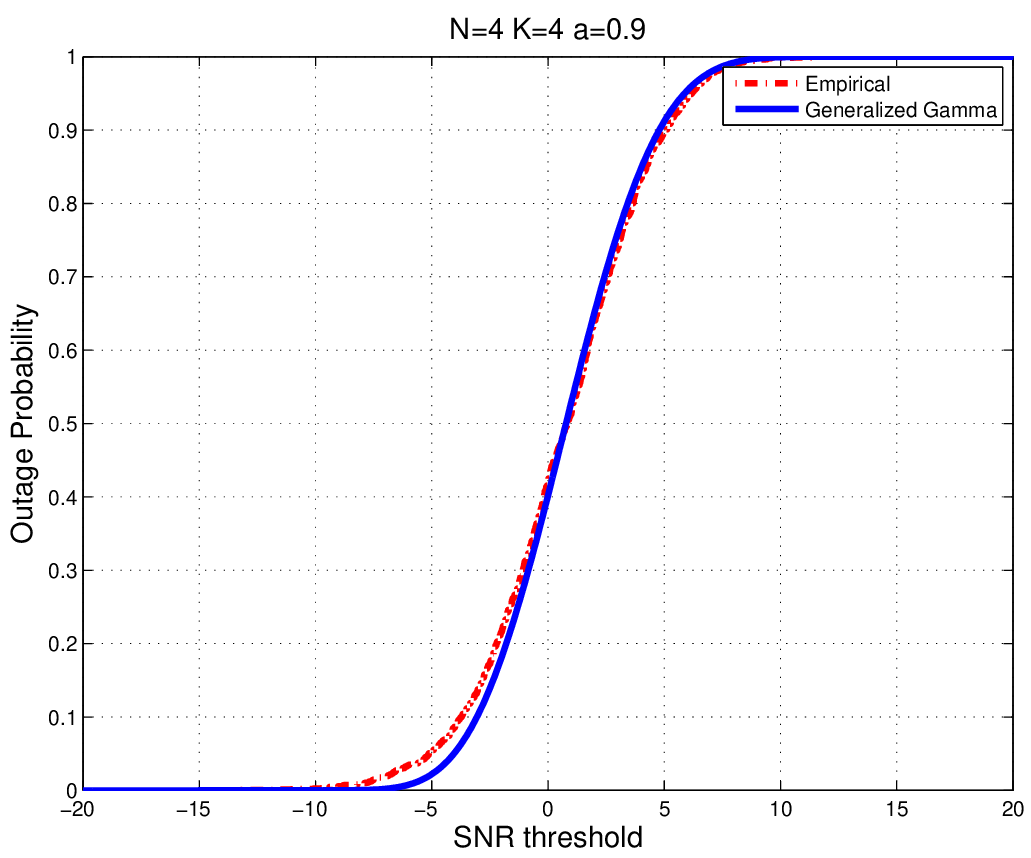}
}
\centering
\subfigure[$N=2K=4$ and $a=0$]{
\includegraphics[width=0.4\textwidth]{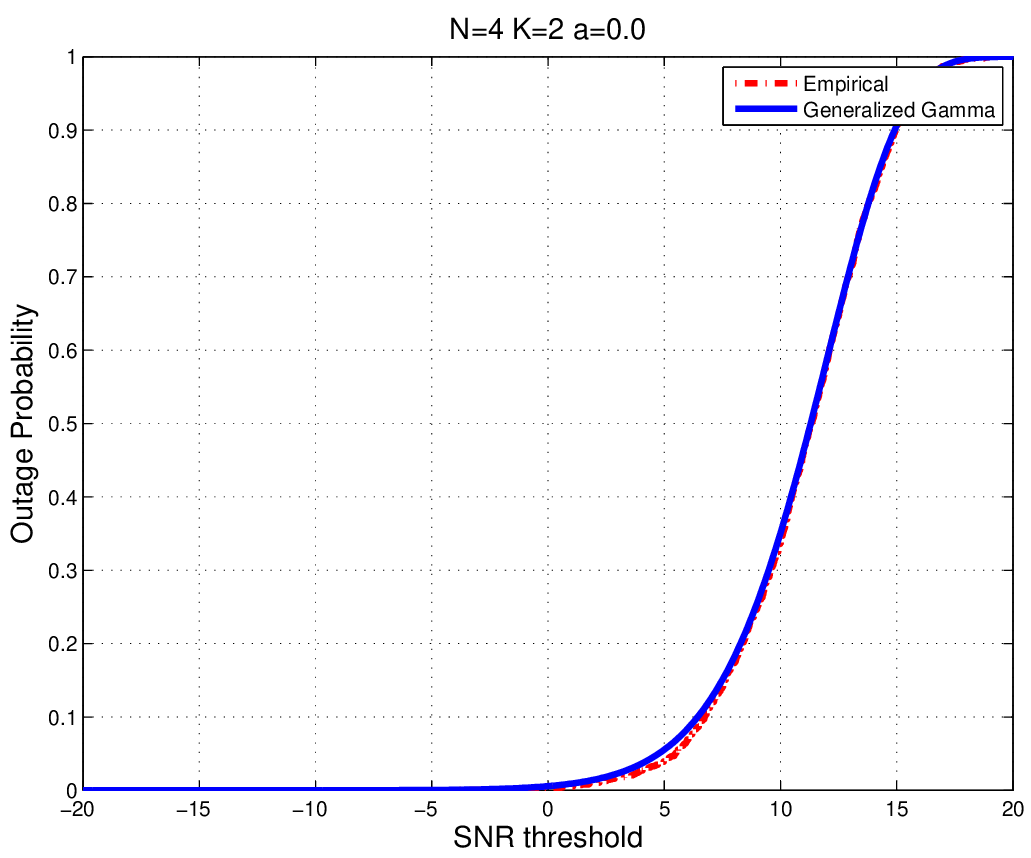}
}
\subfigure[$N=2K=4$ and $a=0.9$]{
\includegraphics[width=0.4\textwidth]{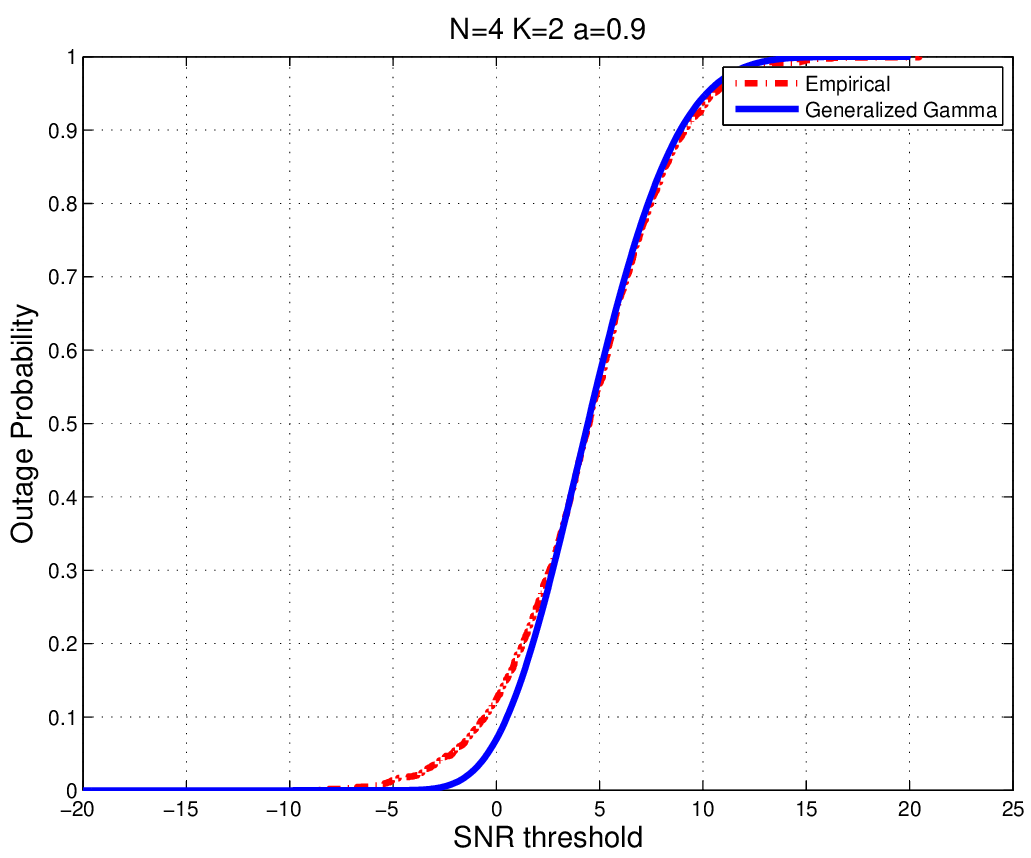}
}
\caption{Outage Probability vs SNR threshold}
\label{fig-outage} 
\end{figure*}

In our simulations, we consider a MIMO system in the uplink direction. The base station is equipped  with $N$ receiving antennas and detects the symbols transmitted by a particular user in the presence of $K$  interfering users. We assume that the correlation matrix ${\bf \Psi}$ is given by ${\bf \Psi}(i,j)=\sqrt{\frac{K}{N}} a^{|i-j|}$ with $0\leq a < 1$.
Recall that $\widetilde{\bf P}$ is the matrix of the interfering users' powers.
We set $\widetilde{\bf P}$ (up to a permutation of its diagonal elements) to:
$$\widetilde{\bf P} = \left\{
\begin{array}{l}
\text{diag}(\left[4P \ \  5P\right])  \ \ \ \mathrm{if}  \ \ K=2 \\
\text{diag}(\left[P\ \  P\ \  2P\ \  4P\right])  \ \ \ \mathrm{if} \ \ K=4 \\
\end{array}
\right.,
$$
where $P$ is the power of the user of interest.
For $K=2^p$ with $3 \le p \le 5$, we assume that the powers of the
interfering sources are arranged into five classes as in Table \ref{tab-powers}.
\begin{table}[htbp]
\label{tab-powers} 
\caption{Power classes and relative frequencies}
\begin{center}
\begin{tabular}{|c|c|c|c|c|c|}
\hline
Class & 1 &2 & 3 & 4 & 5 \\
\hline
Power & $P$ & $2P$ & $4P$ & $8P$ & $16P$ \\
\hline
Relative frequency & 
${1} / {8}$ & ${1} / {4}$ & ${1} / {4}$ & ${1} / {8}$ & ${1} / {4}$\\
\hline
\end{tabular}
\end{center}
\end{table}
We investigate the impact of the correlation coefficient $a$ on the accuracy 
of the asymptotic moments when the input SNR is set to $15$dB for 
$N=K$ (Fig. \ref{fig:rel_err}) and $N=2K$ (Fig. \ref{fig:rel_err_neq}).  
In these figures, the relative error on the estimated first three moments $\frac{\left|\mu_{\infty}-\mu\right|}{\mu}$ ( $\mu_{\infty}$ and $\mu$ denote respectively the asymptotic and empirical moment )
is depicted with respect to the correlation coefficient $a$. These simulations
show that when the number of antennas is small, the asymptotic approximation 
of the second and third moments degrades for large correlation coefficients 
($a$ close to one). 
Despite these discrepancies for $a$ close to $1$, simulations show that 
the BER and the outage probability are well approximated even for small
system dimensions. 
Indeed, Figure \ref{fig:BER-vs-snr} shows the evolution of the empirical BER and
the theoretical BER predicted by (\ref{eq:ber}) versus the input SNR for
different values of $a$, $K$ and $N$. In Figure \ref{fig-outage}, the saddle 
point approximate of the outage probability given by (\ref{eq:saddle}) is
compared with the empirical one.
In both Figures \ref{fig:BER-vs-snr} and \ref{fig-outage}, $2000$ channel 
realizations have been considered, and in Fig. \ref{fig-outage}, the input
SNR has been set to $15$ dB. 
These figures show that even for small system dimensions, the BER is well 
approximated for a wide range of SNR values. The outage probability is also 
well approximated except for small values of the SNR threshold that are 
likely to be in the tail of the asymptotic distribution. 

\appendices
\section{Proof of Theorem \ref{asymptotic}}
\label{app:proof}
In the sequel, we shall heavily rely on the results and techniques
developed in \cite{capacity_analysis}. In the sequel, ${\bf D}$ and
$\widetilde{\bf D}$ are respectively $N\times N$ and $K\times K$
diagonal matrices which satisfy \ref{ass:first} and \ref{ass:second},
${\bf Z}$ is a $N\times K$ matrix whose entries are i.i.d. standard
complex Gaussian, ${\bf X}$ is a $N\times K$ matrix defined by:
$$
{\bf X}={\bf D}^{\frac{1}{2}}{\bf Z}\widetilde{\bf D}^{\frac{1}{2}}\ .
$$
We shall often write ${\bf X}= [{\bf x}_1,\cdots, {\bf x}_K]$ where
the ${\bf x}_j$'s are ${\bf X}$'s columns. We recall hereafter the
mathematical tools that will be of constant use in the sequel.
\subsection{ Notations}
Define the resolvant matrix ${\bf H}$ by:
$$
{\bf H}=\left(\frac{t}{K}{\bf D }^{\frac{1}{2}}{\bf Z}\widetilde{\bf
    D}{\bf Z}^*{\bf D }^{\frac{1}{2}}+{\bf I}_N\right)^{-1}=
\left(\frac{t}{K}{\bf X}{\bf X}^*+{\bf I}_N\right)^{-1}\ .
$$
We introduce the following intermediate quantities:
$$
\beta(t)=\frac{1}{K}\mathrm{Tr}({\bf DH}), \quad \alpha(t)=\frac{1}{K}\mathrm{Tr}({\bf D}\mathbb{E}{\bf H})
\quad \textrm{and} \quad \stackrel{o}{\beta}=\beta-\alpha\ .
$$
Matrix $\widetilde{\bf
  R}(t)=\mathrm{diag}\left(\tilde{r}_1,\cdots,\tilde{r}_K\right)$ is a $K\times K$ diagonal matrix defined by:
$$\widetilde{\bf R}(t)=\left({\bf I}+t\alpha(t)\widetilde{\bf D}_K\right)^{-1}\ .$$
Let $\tilde{\alpha}=\frac{1}{K}\mathrm{Tr}(\widetilde{\bf
  D}\widetilde{\bf R})$. Then, matrix ${\bf
  R}(t)=\mathrm{diag}\left({r}_1,\cdots,{r}_N\right)$ is a $N\times N$ matrix defined by:
$$
{\bf R}(t)=\left({\bf I}+t\tilde{\alpha}(t){\bf D}\right)^{-1}.
$$

\subsection{Mathematical Tools} 
The results below, of constant use in the proof of Theorem
\ref{asymptotic}, can be found in \cite{capacity_analysis}.
\subsubsection{Differentiation formulas }
\begin{eqnarray}
  \frac{\partial H_{pq}}{\partial{X_{ij}}}&=&-\frac{t}{K}\left[{\bf X}^*{\bf H}\right]_{jq}{H}_{pi}=-\frac{t}{K}\left[{\bf x}_j^*{\bf H}\right]_qH_{pi}.\label{eq:diffx}\\
  \frac{\partial H_{pq}}{\partial{\overline{X_{ij}}}}&=&-\frac{t}{K}\left[{\bf HX}\right]_{pj}H_{iq}=-\frac{t}{K}\left[{\bf Hx}_j\right]_pH_{iq}\label{eq:diffxbar}\\
  \nonumber
\end{eqnarray}
\subsubsection{Integration by parts formula for Gaussian functionals}: Let ${ \Phi}$
  be a $\mathcal{C}^1$ complex function polynomially bounded together
  with its derivatives, then:
\begin{equation}
\mathbb{E}\left[X_{ij}{\Phi}({\bf X})\right]=d_i\tilde{d}_j\mathbb{E}\left[\frac{\partial
 { \Phi}({\bf X})}{\partial\overline{X_{ij}}}\right].
\label{eq:derivation}
\end{equation}
\subsubsection{Poincar\'{e}-Nash inequality}
Let ${\bf X}$ and ${{\Phi}}$ be as above, then:
\begin{equation}
  \mathrm{Var}({ \Phi}({\bf X}))
\leq \sum_{i=1}^N\sum_{j=1}^K d_i\tilde{d}_j 
\mathbb{E}\left[\left|\frac{\partial{\Phi}({\bf X})}{\partial X_{ij}}
\right|^2 + \left|\frac{\partial{\Phi}({\bf X})}{\partial \overline{X_{ij}}}\right|^2\right].
\label{poincare}
\end{equation}

\subsubsection{Deterministic approximations and various estimations}
\begin{proposition}
\label{prop:RT}
Let $\left({\bf A}_K\right)$ and $\left({\bf B}_K\right)$ be two
sequences of respectively $N\times N$ and $K\times K$ diagonal
deterministic matrices whose spectral norm are uniformly bounded in
$K$, then the following hold true:
$$
\frac{1}{K}\mathrm{Tr}({\bf AR})=\frac{1}{K}\mathrm{Tr}({\bf
  AT})+\mathcal{O}(K^{-2}),\qquad
\frac{1}{K}\mathrm{Tr}({\bf B}\widetilde{\bf R})=\frac{1}{K}\mathrm{Tr}({\bf B}\widetilde{\bf T})+\mathcal{O}(K^{-2}).
$$
\end{proposition}
\begin{proposition}
\label{prop:variance}
Let $\left({\bf A}_K\right)$, $\left({\bf B}_K\right)$ and $\left({\bf C}_K\right)$ be three
sequences of $N\times N$, $K\times K$ and $N\times N$ diagonal
deterministic matrices whose spectral norm are uniformly bounded in
$K$. Consider the following functions:
$$
\Phi({\bf X})=\frac{1}{K}\mathrm{Tr}\left({\bf AH}\frac{{\bf
      XBX}^*}{K}\right), \hspace{0.5cm} \Psi({\bf
  X})=\frac{1}{K}\mathrm{Tr}\left({\bf AHDH}\frac{{\bf
      XBX}^*}{K}\right).
$$
Then, 
\begin{enumerate}
\item the following estimations hold true: 
$$
\mathrm{var}\,\Phi({\bf X}),\ \mathrm{var}\,\Psi({\bf X}),\ \mathrm{var}(\beta)\quad \textrm{and}\quad  
\mathrm{var}\left(\frac 1K \mathrm{Tr}{\bf AHCH}\right)\quad \textrm{are}\quad \mathcal{O}(K^{-2})\ .
$$
\item the following approximations hold true:
\begin{eqnarray}
\mathbb{E}\left[\Phi({\bf X})\right]\!\!\!\!&=&\!\!\!\!\frac{1}{K}\mathrm{Tr}\left(\widetilde{\bf D}\widetilde{\bf T}{\bf B}\right)\frac{1}{K}\mathrm{Tr}\left({\bf ADT}\right)+\mathcal{O}(K^{-2}),\label{eq:phi}\\
\mathbb{E}\left[\Psi({\bf X})\right]\!\!\!\!&=&
\!\!\!\!\frac{1}{1-t^2\gamma\tilde{\gamma}}\left(\frac{1}{K^2}\mathrm{Tr}\left(\widetilde{\bf D}\widetilde{\bf T}{\bf B}\right)\mathrm{Tr}({\bf AD}^2{\bf T}^2)-\frac{t\gamma}{K^2}\mathrm{Tr}\left(\widetilde{\bf D}^2\widetilde{\bf T}^2{\bf B}\right)\mathrm{Tr}({\bf ADT})\right)+\mathcal{O}(K^{-2}),\ \ \ \ \ 
\label{eq:Psi}\\
\mathbb{E}\frac{1}{K}\mathrm{Tr}\left[{\bf AHDH}\right]\!\!\!\!&=&\!\!\!\!\frac{1}{1-t^2\gamma\tilde{\gamma}}\frac{1}{K}\mathrm{Tr}({\bf AD}{\bf T}^2)+\mathcal{O}(K^{-2}).
\label{eq:ADT}\\
\nonumber
\end{eqnarray}
\end{enumerate}
\end{proposition}
Proofs of Propositions \ref{prop:RT} and \ref{prop:variance} are
essentially provided in \cite{capacity_analysis}. In the same vein, the
following proposition will be needed.
\begin{proposition}
\label{prop:variance_new}
Let $\left({\bf A}_K\right)$, $\left({\bf B}_K\right)$ and $\left({\bf C}_K\right)$ be three
sequences of $N\times N$, $K\times K$ and $N\times N$ diagonal
deterministic matrices whose spectral norm are uniformly bounded in
$K$. Consider the following function:
$$
\varphi({\bf X})=\frac{1}{K}\mathrm{Tr}\left[{\bf CHAHAH}\frac{{\bf XBX}^*}{K}\right]\ .
$$
Then
$
\mathrm{var}\,\varphi({\bf X})=\mathcal{O}(K^{-2})$ and 
$\mathrm{var}\left(\frac{1}{K}\mathrm{Tr}{\bf AHAHAH}\right)=\mathcal{O}(K^{-2})\ .
$
\end{proposition}
Proof of Proposition \ref{prop:variance_new} is essentially the same
as the proof of Proposition \ref{prop:variance}-1). It is provided for
completeness and postponed to appendix \ref{appendix-deux}.
\subsection{End of proof of Theorem \ref{asymptotic}}

We are now in position to complete the proof of Theorem \ref{asymptotic}. 
Using the notations of \cite{capacity_analysis}, the SNR writes:
$$\beta_K=\frac{tp_0}{K}{\bf z}^*{\bf D}^{\frac{1}{2}}{\bf H(t)}{\bf D}^{\frac{1}{2}}{\bf z},$$
where $t=\frac{1}{\rho}$. Hence, the third moment is given by:
\begin{eqnarray}
\mathbb{E}\left(\beta_K-\mathbb{E}\beta_K\right)^3&=&\frac{(tp_0)^3}{K^3}\mathbb{E}\left({\bf z}^*{\bf D}^{\frac{1}{2}}{\bf H}{\bf D}^{\frac{1}{2}}{\bf z}-\mathbb{E}\mathrm{Tr}{\bf DH}\right)^3,\nonumber\\
&=&\frac{(tp_0)^3}{K^3}\mathbb{E}\left({\bf z}^*{\bf D}^{\frac{1}{2}}{\bf H}{\bf D}^{\frac{1}{2}}{\bf z}-\mathrm{Tr}{\bf DH}+\mathrm{Tr}{\bf DH}-\mathbb{E}\mathrm{Tr}{\bf DH}\right)^3,\nonumber\\
&=&\frac{(tp_0)^3}{K^3}\left[\mathbb{E}\left({\bf z}^*{\bf D}^{\frac{1}{2}}{\bf H}{\bf D}^{\frac{1}{2}}{\bf z}-\mathrm{Tr}{\bf DH}\right)^3+3\mathbb{E}\left({\bf z}^*{\bf D}^{\frac{1}{2}}{\bf H}{\bf D}^{\frac{1}{2}}{\bf z}-\mathrm{Tr}{\bf DH}\right)^2\left(\mathrm{Tr}{\bf DH}-\mathbb{E}\mathrm{Tr}{\bf DH}\right)\right.\nonumber\\
&&\left.+3\mathbb{E}\left({\bf z}^*{\bf D}^{\frac{1}{2}}{\bf H}{\bf D}^{\frac{1}{2}}{\bf z}-\mathrm{Tr}{\bf DH}\right)\left(\mathrm{Tr}{\bf DH}-\mathbb{E}\mathrm{Tr}{\bf DH}\right)^2+\mathbb{E}\left(\mathrm{Tr}{\bf DH}-\mathbb{E}\mathrm{Tr}{\bf DH}\right)^3\right],\nonumber\\
&=&\frac{(tp_0)^3}{K^3}\left[\mathbb{E}\left({\bf z}^*{\bf D}^{\frac{1}{2}}{\bf H}{\bf D}^{\frac{1}{2}}{\bf z}-\mathrm{Tr}{\bf DH}\right)^3+3\mathbb{E}\left({\bf z}^*{\bf D}^{\frac{1}{2}}{\bf H}{\bf D}^{\frac{1}{2}}{\bf z}-\mathrm{Tr}{\bf DH}\right)^2\left(\mathrm{Tr}{\bf DH}-\mathbb{E}\mathrm{Tr}{\bf DH}\right)\right.\nonumber\\
&&\left.+\mathbb{E}\left(\mathrm{Tr}{\bf DH}-\mathbb{E}\mathrm{Tr}{\bf DH}\right)^3\right]\label{eq:last1}
\end{eqnarray}
In order to deal with the first term of the right-hand side of (\ref{eq:last1}), notice that
if ${\bf M}$ is a deterministic matrix and ${\bf x}$ is a standard Gaussian vector, then:
$$
\mathbb{E}\left({\bf x}^*{\bf M}{\bf x}-\mathrm{Tr}{\bf
    M}\right)^3=\mathrm{Tr}({\bf
  M}^3)\mathbb{E}\left(|x_1|^2-1\right)^3
$$
(such an identity can be easily proved by considering the spectral
decomposition of ${\bf M}$). Hence,
\begin{eqnarray*}
\mathbb{E}\left({\bf z}^*{\bf D}^{\frac{1}{2}}{\bf H}{\bf D}^{\frac{1}{2}}{\bf z}-\mathrm{Tr}{\bf DH}\right)^3&=&\mathbb{E}\mathrm{Tr} \left({\bf DH}\right)^3\mathbb{E}\left(|Z_{11}|^2-1\right)^3,\\
&=&2\mathbb{E}\mathrm{Tr}\left({\bf DHDHDH}\right).
\end{eqnarray*}
The second term of the right-hand side of (\ref{eq:last1}) is uniformly bounded in $K$. Indeed:
\begin{eqnarray*}
  3\mathbb{E}\left({\bf z}^*{\bf D}^{\frac{1}{2}}{\bf H}{\bf D}^{\frac{1}{2}}{\bf z}-\mathrm{Tr}({\bf DH})\right)^2&=&3\mathbb{E}\left(|Z_{11}|^2-1\right)^2\mathrm{Tr}{\bf DHDH}\left(\mathrm{Tr}{\bf DH}-\mathbb{E}\mathrm{Tr}{\bf DH}\right),\\
  &\leq& 3\sqrt{\mathrm{var}\left(\mathrm{Tr}{\bf DHDH}\right)}\sqrt{\mathrm{var}\left(\mathrm{Tr}{\bf DH}\right)}
\end{eqnarray*}
which is ${\mathcal O}(1)$ according to Proposition
\ref{prop:variance}.  It remains to deal with
$\mathbb{E}\left(\mathrm{Tr}{\bf DH}-\mathbb{E}\mathrm{Tr}{\bf
    DH}\right)^3$, which can be proved to be uniformly bounded in $K$
using concentration results for the spectral measure of random
matrices \cite{zeitou} (see also \cite[eq.(86)-(87)]{cioffi}, where
details are provided). Consequently, we end up with the following
approximation:
$$
K^2\mathbb{E}\left(\beta_K-\mathbb{E}\beta_K\right)^3=\frac{(tp_0)^3}{K}\mathbb{E}\left(|Z_{11}|^2-1\right)^3\mathbb{E}\mathrm{Tr}{\bf
  DHDHDH}+\mathcal{O}\left(K^{-1}\right)
$$
which is deterministic but still depends on the distribution of the
entries via the expectation operator $\mathbb{E}$. The rest of the proof is
devoted to provide a deterministic approximation of
$\mathbb{E}\mathrm{Tr}\left({\bf DHDHDH}\right)$ depending on $\gamma$,
$\tilde \gamma$, ${\bf T}$ and $\widetilde{\bf T}$.

Note that ${\bf H}={\bf I}-\frac{t}{K}{Â \bf HXX}^*$, thus:
\begin{eqnarray}
\left[{\bf HDHDH}\right]_{pp}&=&\left[{\bf HDHD}\right]_{pp}-t\left[{\bf HDHDH}\frac{\bf XX^*}{K}\right]_{pp},\nonumber\\
&=&\left[{\bf HDHD}\right]_{pp}-\frac{t}{K}\sum_{j=1}^K \left[{\bf HDHDHx}_j\right]_p\overline{{ X}_{pj}}.\label{eq:identity}\\
\nonumber
\end{eqnarray}
Let us deal with the second term of (\ref{eq:identity}). We have:
$$
\mathbb{E}\frac{1}{K} \left[{\bf HDHDHx}_j\right]_p\overline{{ X}_{pj}}=\frac{1}{K}\sum_{k=1}^N\mathbb{E}\left(\left[{\bf HDHDH}\right]_{pk}X_{kj}\overline{X_{pj}}\right).
$$
Using the integration by part formula (\ref{eq:derivation}), we get:
\begin{eqnarray*}
\mathbb{E}\left[{\bf HDHDHx}_j\right]_p\overline{{ X}_{pj}}
&=& \sum_{k=1}^N d_k\tilde{d}_j  \delta(p-k) 
\mathbb{E} \left[{\bf HDHDH}\right]_{pk} +  \sum_{k=1}^N d_k \tilde{d}_j 
\mathbb{E}\left[\overline{X_{pj}} \sum_{\ell,m=1}^{N}\frac{\partial \left[H_{p\ell}d_{\ell}d_m H_{\ell m}  H_{mk}\right]}
{\partial{\overline{X_{kj}}}}\right],\\
&=& d_p \tilde{d}_j 
\mathbb{E} \left[{\bf HDHDH}\right]_{pp}-\frac{t}{K}
\sum_{k,\ell,m=1}^N d_k \tilde{d}_j d_m d_{\ell} 
\mathbb{E}\left[\overline{X_{pj}} \left[{\bf Hx}_j\right]_p H_{k\ell}H_{\ell m}H_{mk}\right] \\
&&-\frac{t}{K}\sum_{k,\ell,m=1}^N d_k \tilde{d}_j d_m d_{\ell} 
\mathbb{E}\left[\overline{X_{pj}}H_{p\ell} \left[{\bf Hx}_j\right]_{\ell} H_{km}H_{mk}\right]\\
&&-\frac{t}{K}\sum_{k,\ell,m=1}^N d_k \tilde{d}_j d_m d_{\ell}
\mathbb{E}\left[ H_{p\ell} H_{\ell m} \left[{\bf Hx}_j\right]_m H_{kk}\right].\\
&=& d_p \tilde{d}_j 
\mathbb{E} \left[{\bf HDHDH}\right]_{pp}-\frac{t}{K}\tilde{d}_j \mathbb{E}\left[\left[{\bf Hx}_j\right]_p
\overline{X_{pj}}\mathrm{Tr}\left({\bf DHDHDH}\right)\right] \\
&&-\frac{t}{K}\tilde{d}_j \mathbb{E}\left[\left[{\bf HDHx}_j\right]_p\overline{X_{pj}}\mathrm{Tr}
\left({\bf DHDH}\right)\right] -
\frac{t}{K}\tilde{d}_j\mathbb{E}\left[\left[{\bf HDHDHx}_j\right]_p\overline{X_{pj}}\mathrm{Tr}\left({\bf DH}
\right)\right].\\
\end{eqnarray*}

Substituting in the last term $\frac{1}{K}\mathrm{Tr}{\bf
  DH}=\stackrel{o}{\beta}+\alpha$ where
$\stackrel{o}{\beta}=\beta-\alpha$, we get:

\begin{eqnarray*}
\mathbb{E}\left[{\bf HDHDHx}_j\right]_p\overline{{ X}_{pj}}&=& 
d_p \tilde{d}_j \mathbb{E} \left[{\bf HDHDH}\right]_{pp}
-\frac{t}{K}\tilde{d}_j \mathbb{E}\left[\left[{\bf Hx}_j\right]_p
\overline{X_{pj}}\mathrm{Tr}\left({\bf DHDHDH}\right)\right] \\
&&-\frac{t}{K}\tilde{d}_j \mathbb{E}\left[\left[{\bf HDHx}_j\right]_p
\overline{X_{pj}}\mathrm{Tr}\left({\bf DHDH}\right)\right] 
-t\tilde{d}_j\mathbb{E}\left[\left[{\bf HDHDHx}_j\right]_p\overline{X_{pj}}\stackrel{o}{\beta}\right]\\
&&-t\tilde{d}_j\mathbb{E}\left[\left[{\bf HDHDHx}_j\right]_p\overline{X_{pj}}\right]\alpha.
\end{eqnarray*}

Therefore, we have:
\begin{eqnarray*}
  \left(1+t\alpha\tilde{d}_j\right)\mathbb{E}\left[\left[{\bf HDHDHx}_j\right]_p \overline{X_{pj}}\right] &=&
  d_p \tilde{d}_j \mathbb{E}\left[{\bf HDHDH}\right]_{pp}
  -\frac{t}{K}\mathbb{E}\left[\left[{\bf Hx}_j\right]_p
    \overline{X_{pj}}\tilde{d}_j\mathrm{Tr}\left[{\bf DHDHDH}\right]\right]\\
  &&-\frac{t}{K}\tilde{d}_j \mathbb{E}\left[\left[{\bf HDHx}_j\right]_p \overline{X_{pj}}\mathrm{Tr} 
    \left[{\bf DHDH}\right]\right]-t\tilde{d}_j
\mathbb{E}\left[\left[{\bf HDHDHx}_j\right]_p \overline{X_{pj}}\stackrel{o}{\beta}\right]. 
\end{eqnarray*}

Multiplying the right hand and the left hand sides by  $\tilde{r}_j=\frac{1}{1+t\alpha\tilde{d}_j}$, we get:
\begin{multline}
\mathbb{E} \left[{\bf HDHDHx}_j\right]_p \overline{X_{pj}}\quad=\quad
\tilde{r}_jd_p\tilde{d}_j\mathbb{E}\left[{ \bf HDHDH}\right]_{pp}
-\frac{t}{K}\tilde{r}_j\mathbb{E}\left[\left[{\bf Hx}_j\right]_p\overline{X_{pj}}\tilde{d}_j
\mathrm{Tr}\left[{\bf DHDHDH}\right]\right]\\
-\frac{t}{K}\tilde{d}_j\tilde{r}_j\mathbb{E}\left[\left[{\bf HDHx}_j\right]_p\overline{X_{pj}}
\mathrm{Tr}\left[{\bf DHDH}\right]\right]-t\tilde{d}_j\tilde{r}_j
\mathbb{E}\left[\left[{\bf HDHDHx}_j\right]_p\overline{X_{pj}}\stackrel{o}{\beta}\right].
\label{eq:previous}
\end{multline}
Plugging  (\ref{eq:previous}) into (\ref{eq:identity}), we obtain:
\begin{eqnarray*}
  \mathbb{E}\left[{\bf HDHDH}\right]_{pp}&=&\mathbb{E}\left[{\bf HDHD}\right]_{pp}
  -\sum_{j=1}^K \frac{t}{K}\tilde{r}_j d_p 
  \tilde{d}_j\mathbb{E}\left[{\bf HDHDH}\right]_{pp}+\frac{t^2}{K^2}\sum_{j=1}^{K}\tilde{r}_j
  \mathbb{E}\left[{\bf Hx}_j\right]_p\overline{X_{pj}}\tilde{d}_j\mathrm{Tr}\left[{\bf DHDHDH}\right]\\
  &&+\frac{t^2}{K^2}\sum_{j=1}^K \tilde{d}_j\tilde{r}_j
  \mathbb{E}\left[{\bf HDHx}_j\right]_p\overline{X_{p,j}}\mathrm{Tr}\left[{\bf DHDH}\right]
  +\frac{t}{K}\sum_{j=1}^K \tilde{d}_j\tilde{r}_j
  \mathbb{E}\left[{\bf HDHDHx}_j\right]_p\overline{X_{p,j}}\stackrel{o}{\beta},\\
  &=&\mathbb{E}\left[{\bf HDHD}\right]_{pp}
  -t\tilde{\alpha}d_p\mathbb{E}\left[{\bf HDHDH}\right]_{pp}
  +\frac{t^2}{K^2}\mathbb{E}\mathrm{Tr}({\bf DHDHDH})
\left[{\bf HX\widetilde{\bf R}\widetilde{\bf D}{\bf X^*}}\right]_{pp}\\
  &&+\frac{t^2}{K^2}\mathbb{E}\mathrm{Tr}\left[{\bf DHDH}\right]
\left[{\bf HDHX}\widetilde{\bf D}{\widetilde{\bf R}{\bf X^*}}\right]_{pp}
+\frac{t^2}{K}\mathbb{E}\stackrel{o}{\beta}\left[{\bf HDHDHX\widetilde{\bf D}\widetilde{\bf R}{X}^*}\right]_{pp}.
\end{eqnarray*}
Hence,
\begin{eqnarray*}
(1+t\tilde{\alpha}d_p)\mathbb{E}\left[{\bf HDHDH}\right]_{pp}&=&\mathbb{E}\left[{\bf HDHD}\right]_{pp}
+\frac{t^2}{K^2}\mathbb{E}\mathrm{Tr}\left[{\bf DHDHDH}\right]\left[{\bf HX}
\widetilde{\bf R}\widetilde{\bf D}{\bf X}^*\right]_{pp}\\
&&+\frac{t^2}{K^2}\mathbb{E}\mathrm{Tr}\left[{\bf DHDH}\right]
\left[{\bf HDHX}\widetilde{\bf D}\widetilde{\bf R}{\bf X^*}\right]_{pp}
+\frac{t^2}{K}\mathbb{E}\stackrel{o}{\beta}\left[{\bf HDHDHX\widetilde{D}\widetilde{R}X}^*\right]_{pp}\ .
\end{eqnarray*}
Multiplying the left and right hand sides by $r_p=\frac{1}{1+t\tilde{\alpha}d_p}$, we get:
\begin{eqnarray}
\mathbb{E}\left[{\bf HDHDH}\right]_{pp}&=&r_p\mathbb{E}\left[{\bf HDHD}\right]_{pp}
+\frac{t^2}{K^2}r_p\mathbb{E}\mathrm{Tr}\left[{\bf DHDHDH}\right]\left[{\bf HX}
\widetilde{\bf R}\widetilde{\bf D}{\bf X}^*\right]_{pp}\nonumber\\
&&+\frac{t^2}{K^2}r_p\mathbb{E}\mathrm{Tr}\left[{\bf DHDH}\right]\left[{\bf HDHX}
\widetilde{\bf D}\widetilde{\bf R}{\bf X}^*\right]_{pp}
+\frac{t^2}{K}r_p\mathbb{E}\stackrel{o}{\beta}\left[{\bf HDHDHX}\widetilde{\bf D}\widetilde{\bf R}{\bf X}^*\right]_{pp}.
\label{eq:last}
\end{eqnarray}
Multiplying by $d_p$, summing over $p$ and dividing by $K$, we obtain:
\begin{eqnarray}
\mathbb{E}\frac{1}{K}\mathrm{Tr}\left[{\bf DHDHDH}\right]
&=&\mathbb{E}\frac{1}{K}\sum_{p=1}^K d_p\left[{\bf HDHDH}\right]_{pp},\nonumber\\
&=&\frac{1}{K}\sum_{p=1}^K r_p d_p \mathbb{E}\left[{\bf HDHD}\right]_{pp}
+\frac{t^2}{K^3}\mathbb{E}\, \mathrm{Tr}\left({\bf DHDHDH}\right)
\mathrm{Tr}\left({\bf DRH}{\bf X\widetilde{\bf R}\widetilde{\bf D}{\bf X}^*}\right)\nonumber\\
&&+\frac{t^2}{K^3}\mathbb{E}\mathrm{Tr}\left({\bf DHDH}\right)
\mathrm{Tr}\left({\bf DRHDHX}\widetilde{\bf D}\widetilde{\bf R}{\bf X}^*\right)\nonumber\\
&&+\frac{t^2}{K^2}\mathbb{E}\ \stackrel{o}{\beta}
\mathrm{Tr}\left({\bf DRHDHDH}{\bf X\widetilde{D}\widetilde{R}X^*}\right),\nonumber\\
&\stackrel{\triangle}=&\chi_1+\chi_2+\chi_3+\chi_4,\label{eq:sum}
\end{eqnarray}
where:
\begin{eqnarray*}
\chi_1&=&\frac{1}{K}\mathbb{E}\mathrm{Tr}\left({\bf DRHDHD}\right)\ ,\\
\chi_2&=&\frac{t^2}{K}\mathbb{E}
\mathrm{Tr}\left({\bf DHDHDH}\right)
\frac{1}{K}\mathrm{Tr}\left({\bf DRH}\frac{\bf X\widetilde{D}\widetilde{R}X^*}{K}\right)\ ,\\
\chi_3&=&\frac{t^2}{K}\mathbb{E}\mathrm{Tr}\left({\bf DHDH}\right)
\frac{1}{K}\mathrm{Tr}\left({\bf DRHDH}\frac{\bf X\widetilde{D}\widetilde{R}X^*}{K}\right)\ ,\\
\chi_4&=&\frac{t^2}{K}\mathbb{E}\, \stackrel{o}{\beta}\mathrm{Tr}
\left({\bf DRHDHDH}\frac{\bf X\widetilde{D}\widetilde{R}X^*}{K}\right)\ .
\end{eqnarray*}
According to Proposition \ref{prop:variance}, 
$\mathrm{var}\frac{1}{K}\mathrm{Tr}\left({\bf DRHDHDH}\frac{{\bf X\widetilde{\bf D}\widetilde{\bf R}\bf X}^*}{K}\right)$
is of order $\mathcal{O}(K^{-2})$. Similarly, $\mathrm{var}({\beta})=\mathcal{O}(K^{-2})$. Hence, using Cauchy-Schwartz 
inequality, we get the estimation $\chi_4=\mathcal{O}(K^{-2})$.
It remains to work out the expressions involved in $\chi_1$, $\chi_2$ and $\chi_3$
by removing the terms with expectation and replacing them with deterministic equivalents.

Since $\mathrm{var}\frac{1}{K}\mathrm{Tr}\left({\bf DRH}\frac{{\bf X
      \widetilde{\bf D}\widetilde{\bf R}{\bf X}}^*}{K} \right)
=\mathcal{O}(K^{-2})$ by Proposition \ref{prop:variance} and
$\mathrm{var}(\frac{1}{K}\mathrm{Tr}{\bf DHDHDH})=\mathcal{O}(K^{-2})$
by Proposition \ref{prop:variance_new}, we have:
\begin{eqnarray}
\chi_2&=&\frac{t^2}{K}\mathbb{E}\mathrm{Tr}\left({\bf DHDHDH}\right)
\mathbb{E}\left(\frac{1}{K}\mathrm{Tr}
\left[{\bf DRH}\frac{\bf X\widetilde{\bf D}{\widetilde R}{\bf X}^*}{K}\right]\right)+\mathcal{O}(K^{-2}),\nonumber\\
&\stackrel{(a)}{=}&\frac{t^2}{K}\mathbb{E}\mathrm{Tr}\left({\bf DHDHDH}\right)
\frac{1}{K}\mathrm{Tr}\left(\widetilde{\bf D}\widetilde{\bf T}
\widetilde{\bf D}\widetilde{\bf R}\right)\frac{1}{K}\mathrm{Tr}\left({\bf DRDT}\right)
+\mathcal{O}(K^{-2}),\nonumber\\
&\stackrel{(b)}{=}&\frac{t^2}{K}\mathbb{E}\mathrm{Tr}\left({\bf DHDHDH}\right)
\gamma\tilde{\gamma}+\mathcal{O}(K^{-2})\label{eq:chi_2}\ .
\end{eqnarray}
where (a) follows from Proposition \ref{prop:variance}-2) and (b), from Proposition \ref{prop:RT}.
Similar arguments yield:
\begin{eqnarray}
  \chi_3&=&\frac{t^2}{K}\mathbb{E}\mathrm{Tr}\left({\bf DHDH}\right)
  \mathbb{E}\left(\frac{1}{K}\mathrm{Tr}
    \left[{\bf DRHDH\frac{{\bf X\widetilde{\bf D}\widetilde{\bf R}{\bf X}^*}}{K}}\right]\right)+\mathcal{O}(K^{-2}),\nonumber\\
  &=&\frac{t^2\gamma}{(1-t^2\gamma\tilde{\gamma})^2}\left[\frac{1}{K}\mathrm{Tr}\left(\widetilde{\bf D}\widetilde{\bf T}\widetilde{\bf D}\widetilde{\bf R}\right)\frac{1}{K}\mathrm{Tr}\left({\bf DR}{\bf D}^2{\bf T}^2\right)-\frac{t\gamma}{K}\mathrm{Tr}\left(\widetilde{\bf D}^2\widetilde{\bf T}^2\widetilde{\bf D}\widetilde{\bf R}\right)\frac{1}{K}\mathrm{Tr}({\bf DRDT})\right]+\mathcal{O}(K^{-2})\ ,\nonumber\\
  &=&\frac{t^2\gamma}{(1-t^2\gamma\tilde{\gamma})^2}
  \left[\frac{\tilde\gamma}{K}\mathrm{Tr}({\bf D}^3{\bf T}^3)
  -\frac{t\gamma^2}{K}
  \mathrm{Tr}(\widetilde{\bf D}^3\widetilde{\bf T}^3)\right]
  +\mathcal{O}(K^{-2})\label{eq:chi_3}
\end{eqnarray}
and
\begin{eqnarray}
\chi_1&=&\frac{1}{1-t^2\gamma\tilde{\gamma}}\frac{1}{K}\mathrm{Tr}\left({\bf D}^2{\bf R}{\bf DT}^2\right)+\mathcal{O}(K^{-2})\nonumber\\
&=&\frac{1}{1-t^2\gamma\tilde{\gamma}}
\frac{1}{K}\mathrm{Tr}({\bf D}^3{\bf T}^3)+\mathcal{O}(K^{-2}).\label{eq:chi_1}\\
\nonumber
\end{eqnarray}
 Plugging (\ref{eq:chi_3}), (\ref{eq:chi_2}) and (\ref{eq:chi_1}) into (\ref{eq:sum}), we obtain:
$$
\frac{1}{K}\mathbb{E}\mathrm{Tr}({\bf DHDHDH})=\frac{1}{K(1-t^2\gamma\tilde{\gamma})^3}\mathrm{Tr}{\bf D}^3{\bf T}^3-\frac{t^3\gamma^3}{K(1-t^2\gamma\tilde{\gamma})^3}\mathrm{Tr}\widetilde{\bf T}^3\widetilde{\bf D}^3+\mathcal{O}(K^{-2}).\\
$$
Hence,
\begin{eqnarray*}
  K^2\mathbb{E}\left(\frac{\beta_K}{p_0}-\mathbb{E}\frac{\beta_K}{p_0}\right)^3&=&\frac{\rho^3}{K\left(\rho^2-\gamma\tilde{\gamma}\right)^3}\left[\mathrm{Tr}{\bf D}^3{\bf T}^3-\frac{\gamma^3}{\rho^3}\mathrm{Tr}\widetilde{\bf D}^3\widetilde{\bf T}^3\right]\mathbb{E}\left(\left|Z_{11}\right|^2-1\right)^3+\mathcal{O}\left(\frac{1}{K}\right),\\
  &=&\frac{2\rho^3}{K\left(\rho^2-\gamma\tilde{\gamma}\right)^3}\left[\mathrm{Tr}{\bf D}^3{\bf T}^3-\frac{\gamma^3}{\rho^3}\mathrm{Tr}\widetilde{\bf D}^3\widetilde{\bf T}^3\right]+\mathcal{O}\left(\frac{1}{K}\right).
\end{eqnarray*}
The fact that
$\nu_K=\frac{2\rho^3}{K\left(\rho^2-\gamma\tilde{\gamma}\right)^3}
\left[\mathrm{Tr}{\bf D}^3{\bf
    T}^3-\frac{\gamma^3}{\rho^3}\mathrm{Tr}\widetilde{\bf
    D}^3\widetilde{\bf T}^3\right]$ is of order ${\mathcal O}(1)$ is
straightforward and its proof is omitted. Proof of Theorem
\ref{asymptotic} is completed.

\section{Proof of Proposition \ref{prop:variance_new}}\label{appendix-deux}

The proof mainly relies on Poincar\'e-Nash inequality. 
Using the Poincar\'e-Nash inequality, we have:
$$
\mathrm{var}(\varphi({\bf X}))\leq \sum_{i=1}^N\sum_{j=1}^K d_i
\tilde{d}_j \mathbb{E} \left|\frac{\partial\varphi}{\partial
    X_{ij}}\right|^2 +\sum_{i=1}^N\sum_{j=1}^K d_i
\tilde{d}_j\mathbb{E}\left|\frac{\partial\varphi}{\partial
    \overline{X_{ij}}}\right|^2\\ .
$$
We only deal with the first term of the last inequality (the second
term can be handled similarly).  We have $\varphi({\bf
  X})=\frac{1}{K^2}\sum_{p,r,s,t=1}^N\sum_{u=1}^K c_{pp} H_{pr} A_{rr}
H_{rs} A_{ss}H_{st} X_{tu}B_{uu} X^*_{pu}$. After straightforward
calculations using the differentiation formula \eqref{eq:diffx},
we get that:
\begin{eqnarray*}
\frac{\partial\varphi}{\partial X_{ij}}=\phi_{ij}^{(1)}+\phi_{ij}^{(2)}+\phi_{ij}^{(3)}+\phi_{ij}^{(4)},
\end{eqnarray*}
where:
\begin{eqnarray*}
\phi_{ij}^{(1)}&=&-\frac{t}{K^3}\left[{\bf X}^*{\bf HAHAHXBX}^*{\bf CH}\right]_{ji},\quad
\phi_{ij}^{(2)}\quad =\quad-\frac{t}{K^3}\left[{\bf X}^*{\bf HAHXBX}^*{\bf CHAH}\right]_{ji},\\
\phi_{ij}^{(3)}&=&-\frac{t}{K^3}\left[{\bf X}^*{\bf HXBX}^*{\bf CHAHAH}\right]_{ji},\quad
\phi_{ij}^{(4)}\quad=\quad \frac{1}{K^2}\left[{\bf BX}^*{\bf CHAHAH}\right]_{ji}.\\
\end{eqnarray*}
Hence, $\left|\frac{\partial\varphi}{\partial
    X_{ij}}\right|^2\le 4\left(
  \left|\phi_{ij}^{(1)}\right|^2+
\left|\phi_{ij}^{(2)}\right|^2+\left|\phi_{ij}^{(3)}\right|^2
+\left|\phi_{ij}^{(4)}\right|^2\right)$ and 
\begin{eqnarray*}
  \sum_{i=1}^N\sum_{j=1}^K d_i\tilde{d}_j\mathbb{E}\left[\left|\frac{\partial\varphi}{\partial X_{ij}}\right|^2\right]
  &\leq&\frac{4t^2}{K^6}\mathbb{E}\mathrm{Tr}\left({\bf DHCXBX}^*{\bf HAHAH}{\bf X}\widetilde{\bf D}{\bf X}^*{\bf HAHAH}{\bf XBX}^*{\bf CH}\right)\\
  &&+\frac{4t^2}{K^6}\mathbb{E}\mathrm{Tr}\left({\bf DHAHC X B X}^*{\bf HAH X}\widetilde{\bf D}{\bf X}^*{\bf HAHXBX}^*{\bf CHAH}\right)\\
  &&+\frac{4t^2}{K^6}\mathbb{E}\mathrm{Tr}\left({\bf DHAHAHCXBX}^*{\bf HX}\widetilde{\bf D}{\bf X}^*{\bf HXBX}^*{\bf CHAHAH}\right)\\
  &&+\frac{4}{K^4}\mathbb{E}\mathrm{Tr}\left({\bf DHAHAHCXB}\widetilde{\bf D}{\bf B}{\bf X}^*{\bf CHAHAH}\right).
\end{eqnarray*}
We only prove that the first term of the right hand side is of order
$K^{-2}$; the other terms being handled similarly. Using Cauchy-Schwartz
inequality, we get:
\begin{eqnarray*}
4\sum_{i=1}^N\sum_{j=1}^K d_i \tilde{d}_j \mathbb{E}\left|\phi_{ij}^{1}\right|^2&\leq &\frac{4t^2d_{\mathrm{max}}\|{\bf H}\|^2\|{\bf C}\|^2}{K^6}\mathbb{E}\mathrm{Tr}\left(\left({\bf HA}\right)^2{\bf H}{\bf X}\widetilde{\bf D}{\bf X}^*{\bf H}\left({\bf AH}\right)^2\left({\bf XBX}^*\right)^2\right),\\
&\leq&\frac{4t^2}{K^6}d_{\mathrm{max}}\|{\bf H}\|^2\|{\bf C}\|^2\left(\mathbb{E}\mathrm{Tr}\left({\bf HA}\right)^2{\bf H}{\bf X}\widetilde{\bf D}{\bf X}^*{\bf H}\left({\bf AH}\right)^2\left({\bf HA}\right)^2{\bf H}{\bf X}\widetilde{\bf D}{\bf X}^*{\bf H}\left({\bf AH}\right)^2\right)^{\frac{1}{2}}\\
&\times&\left(\mathbb{E}\mathrm{Tr}\left({\bf XBX}^*\right)^4\right)^{\frac{1}{2}}\\
&\leq&\frac{4t^2}{K^2}d_{\mathrm{max}}\|{\bf H}\|^8\|{\bf C}\|^2\|{\bf A}\|^4\sqrt{\mathbb{E}\frac{1}{K}\left(\frac{{\bf X}\widetilde{\bf D}{\bf X}^*}{K}\right)^2}\sqrt{\mathbb{E}\frac{1}{K}\left(\frac{{\bf X}{\bf B}{\bf X}^*}{K}\right)^4},
\end{eqnarray*}
where the first inequality follows by using the fact that
$\left|\mathrm{Tr}{\bf AB}\right|\leq \|{\bf B}\|\mathrm{Tr}\left({\bf
    A}\right)$, ${\bf A}$ being hermitian non-negative matrix and the
second follows by applyig twice Cauchy-Schwartz inequalities:
$\mathrm{Tr}\left({\bf AB}\right)\leq \sqrt{\mathrm{Tr}\left({\bf
      AA}^*\right)}\sqrt{\mathrm{Tr}\left({\bf BB}^*\right)}$ and
$\mathbb{E}{XY} \leq \sqrt{\mathbb{E}X^2}\sqrt{\mathbb{E}Y^2}$.  We
end up the proof of the first statement by using the fact that
$\frac{1}{K}\mathbb{E}\left[\frac{1}{K}\mathrm{Tr}\left(\frac{1}{K}{\bf
      X}{\bf B}_K{\bf X}^*\right)^n\right]$ is uniformly bounded in
$K$ whenever ${\bf B}_K$ is a sequence of diagonal matrices with
uniformly bounded spectral norm and $n$ is a given integer.

The second statement follows from the resolvent identity:
$$
\frac{1}{K}\mathrm{Tr}{\bf AHAHAH}=\frac{1}{K}\mathrm{Tr}{\bf AHAHA}-\frac{t}{K}\mathrm{Tr}{\bf AHAHAH{\bf XX}^*}.
$$
According to the first part of the proposition,
$$
\mathrm{var}\left(\frac{1}{K}\mathrm{Tr}{\bf AHAHAH{\bf XX}^*}\right)=\mathcal{O}(K^{-2})\ .
$$
Now, $\mathrm{Tr}{\bf AHAHA}= \mathrm{Tr}{\bf A^2HAH}$ and
$\mathrm{var}\frac{1}{K}\mathrm{Tr}{\bf A^2HAH}=\mathcal{O}(K^{-2})$
by Proposition \ref{prop:variance}-1). Hence, applying inequality
$\mathrm{var}(X+Y)\leq \mathrm{var}(X)+
\mathrm{var}(Y)+2\sqrt{\mathrm{var}(X)\mathrm{var}(Y)}$ yields the
desired result. Proof of Proposition \ref{prop:variance_new} is
completed.

\bibliographystyle{IEEEbib}
\bibliography{mybib}

\newpage 

\newpage

\newpage 

\end{document}